\documentclass[letterpaper, 10pt, journal]{IEEEtran}
\usepackage{amsmath,amsfonts}
\usepackage{algorithm}
\usepackage{algpseudocode}
\usepackage{array}
\usepackage{booktabs}
\usepackage[caption=false,font=footnotesize,labelfont=sf,textfont=sf]{subfig}
\usepackage{textcomp}
\usepackage{stfloats}
\usepackage{url}
\usepackage{verbatim}
\usepackage{graphicx}

\usepackage{siunitx}
\DeclareSIUnit{\Mbps}{Mbps}
\DeclareSIUnit{\bps}{bps}
\DeclareSIUnit{\dBm}{dBm}
\DeclareSIUnit{\fps}{fps}
\DeclareSIUnit{\bytes}{bytes}

\usepackage[
bibstyle=ieee,
citestyle=numeric-comp,
maxnames=1,
minnames=1,
backend=biber,
]{biblatex}
\usepackage{xcolor}
\usepackage{fontspec}
\definecolor{aqua}{RGB}{63, 152, 189}
\definecolor{mybgcolor}{HTML}{b2d6e5}
\definecolor{slide1Freq}{HTML}{3e98bd}
\definecolor{bkoffcol}{HTML}{c69404}
\definecolor{labelfont}{HTML}{90288d}
\definecolor{stasenddata}{HTML}{e2f0d9}
\definecolor{algoap}{HTML}{c00000}
\definecolor{phaseBSR}{HTML}{D9EAD3}
\definecolor{phaseUL}{HTML}{D9D2E9}
\definecolor{phaseACK}{HTML}{EAD1DC}
\definecolor{apBcast}{HTML}{BDD7EE}

\usepackage[abbreviations,nomain,nonumberlist]{glossaries-extra}
\setabbreviationstyle{long-short}   

\newabbreviation{ac}{AC}{access category}
\newabbreviation{aid}{AID}{association ID}
\newabbreviation{aifsn}{AIFSN}{arbitration inter-frame space number}
\newabbreviation{ap}{AP}{access point}
\newabbreviation{ari}{ARI}{access request interval}
\newabbreviation{a2p}{A2P}{Announce to Poll}
\newabbreviation{a2s}{A2S}{Announce to Schedule}

\newabbreviation{ba}{BA}{block acknowledgement}
\newabbreviation{bsr}{BSR}{buffer status report}
\newabbreviation{bsrp}{BSRP}{buffer status report poll}
\newabbreviation{bss}{BSS}{basic service set}

\newabbreviation{cbr}{CBR}{constant bit rate}
\newabbreviation{cmab}{CMAB}{combinatorial multi-armed bandit}
\newabbreviation{cts}{CTS}{combinatorial Thompson Sampling}
\newabbreviation{cucb}{CUCB}{combinatorial upper confidence bound}

\newabbreviation{dl}{DL}{downlink}

\newabbreviation{edca}{EDCA}{enhanced distributed channel access}
\newabbreviation{exp3}{EXP3}{Exponential-weight Algorithm for Exploration and Exploitation}
\newabbreviation{exp3m}{EXP3.M}{Exponential-weight Algorithm for Exploration and Exploitation with Multiple Plays}

\newabbreviation{iot}{IoT}{Internet of Things}
\newabbreviation{iiot}{IIoT}{industrial Internet of Things}

\newabbreviation{mab}{MAB}{multi-armed bandit}
\newabbreviation{mac}{MAC}{medium access control}
\newabbreviation{mcs}{MCS}{modulation and coding scheme}
\newabbreviation{mu-edca}{MU~EDCA}{multi-user enhanced distributed channel access}
\newabbreviation{mtu}{MTU}{maximum transmission unit}

\newabbreviation{obo}{OBO}{OFDMA back-off}
\newabbreviation{ocw}{OCW}{OFDMA contention window}
\newabbreviation{ofdma}{OFDMA}{orthogonal frequency-division multiple access}

\newabbreviation{ra}{RA}{random access}
\newabbreviation{rl}{RL}{reinforcement learning}
\newabbreviation[plural={RUs},longplural={resource units}]{ru}{RU}{resource unit}

\newabbreviation{sa}{SA}{scheduled access}
\newabbreviation{sa-ofdma}{SA-OFDMA}{scheduled access orthogonal frequency-division multiple access}
\newabbreviation{sifs}{SIFS}{short inter-frame space}
\newabbreviation{sta}{STA}{station}

\newabbreviation{tf}{TF}{trigger frame}
\newabbreviation{txop}{TXOP}{transmission opportunity}

\newabbreviation{qos}{QoS}{Quality of Service}

\newabbreviation{ucb}{UCB}{upper confidence bound}
\newabbreviation{udp}{UDP}{User Datagram Protocol}
\newabbreviation{ul}{UL}{uplink}
\newabbreviation{uora}{UORA}{uplink OFDMA-based random access}

\newabbreviation{vo}{VO}{voice}

\newabbreviation{xr}{XR}{extended reality}

\usepackage{cleveref}
\usepackage{pbox}
\usepackage[super]{nth}
\usepackage[normalem]{ulem}
\usepackage[pro]{fontawesome6}
\usepackage{tikz}
\usetikzlibrary{arrows.meta, positioning, calc, backgrounds}

\usepackage[caption=false,font=footnotesize,labelfont=sf,textfont=sf]{subfig}

\crefname{figure}{Fig.}{Figs.}
\Crefname{figure}{Figure}{Figures}

\usepackage{comment}
\graphicspath{{./images}}

\usepackage{todonotes}

\newif\ifdraft
\drafttrue

\ifdraft
    \newcommand{\drafttodo}[1]{%
        \todo[inline, color=red!25]{#1}%
    }
    \newcommand{\idea}[1]{%
        \todo[inline, color=green!30]{#1}%
    }
    \newcommand{\question}[1]{%
        \todo[inline, color=blue!25]{#1}%
    }
    \newcommand{\checkme}[1]{%
        \todo[inline, color=orange!35]{#1}%
    }
\else
    \newcommand{\drafttodo}[1]{}
    \newcommand{\idea}[1]{}
    \newcommand{\question}[1]{}
    \newcommand{\checkme}[1]{}
    \newcommand{\draftinline}[1]{}
\fi

\newcommand{\nthd}[1]{#1\textsuperscript{th}}

\begin{document}

\title{
    Evaluating Polling Schemes for OFDMA Uplink Scheduling in
    Time-Sensitive Wi-Fi Networks
}

\author{
\IEEEauthorblockN{
Douglas Dziedzorm Agbeve$^{1}$,~\IEEEmembership{Graduate Student Member~IEEE},
Ferhat Van Herck$^{2}$,
Andrey Belogaev$^{1}$,
and Jeroen Famaey$^{1}$,~\IEEEmembership{Senior Member~IEEE},
}\\
\IEEEauthorblockA{
$^{1}$\textit{IDLab, University of Antwerp -- imec},
$^{2}$\textit{University of Antwerp}, Antwerp, Belgium\\
\{douglas.agbeve, andrei.belogaev, jeroen.famaey\}@uantwerpen.be,
ferhat.vanherck@student.uantwerpen.be
}
\thanks{}}

\markboth{Journal of \LaTeX\ Class Files,~Vol.~14, No.~8, August~2021}%
{Shell \MakeLowercase{\textit{et al.}}: A Sample Article Using IEEEtran.cls for
IEEE Journals}


\maketitle

\begin{abstract}
    IEEE~802.11ax introduced \gls{ofdma} to improve channel utilization through
    concurrent multi-user transmissions. For effective \gls{ul} scheduling, the
    \gls{ap} requires timely knowledge of buffered traffic at the \glspl{sta}.
    Explicit buffer-status polling provides this information, yet each polling
    round can query only as many \glspl{sta} as there are available \glspl{ru}.
    Existing scheduled-access, random-access, activity-based, and learning-based
    schemes address this buffer-status acquisition problem differently and have
    generally been evaluated in isolation under different traffic and network
    conditions. Consequently, existing studies do not establish how the
    performance of the schemes compare across different time-sensitive
    workloads.
    To address this gap, we evaluate common schemes from the Wi-Fi standard and
    literature under three time-sensitive environments comprising a large-scale
    teleconferencing session, an industrial IoT deployment, and an \gls{xr}
    service. The results show that the relative performance of these schemes is
    highly dependent on the traffic characteristics and offered load. In the
    conferencing environment, activity-based and learning-based polling provide
    the best performance, while the conventional contention-based access
    remains competitive under light offered load. In the industrial scenario,
    scheduled-access is most effective at delivering time-sensitive packets
    within their deadlines under heavy load. Under lighter load, however, the
    conventional contention-based access performs best. In \gls{xr},
    \gls{ofdma}-based schemes achieve high video throughput that conventional
    contention-based access cannot sustain. However, among these schemes,
    activity-based polling achieves the lowest deadline-miss ratios for uplink
    head-tracking traffic at the expense of longer video delay tails. These
    findings highlight the importance of selecting channel access and
    scheduling schemes according to the traffic requirements and operating
    conditions.

\end{abstract}


\glsresetall
\section{Introduction}
\label{sec:intro}

\Gls{edca} is the default contention-based channel access mechanism in Wi-Fi.
Each \gls{sta} contends independently for the medium, and the winner transmits
over the full channel bandwidth. This requires no per-transmission coordination
from the \gls{ap}, but scales poorly. As the network becomes denser, collisions
and back-off consume an increasing share of the airtime, degrading the latency
of time-sensitive applications such as industrial automation and
\gls{xr}~\cite{adame2021time}. The IEEE~802.11ax amendment introduced
\gls{ofdma} to improve channel utilization under such conditions. The channel is
partitioned into \glspl{ru}, allowing the \gls{ap} to serve several \glspl{sta}
in one transmission by assigning each its own \gls{ru}~\cite{80211ax}.

In the \gls{ul}, \gls{ofdma} shifts access control to the \gls{ap}, which
coordinates transmissions using \glspl{tf} to assign \glspl{ru} to individual
\glspl{sta}. The standard also allows contention-based \gls{edca} access to be
suppressed at the \glspl{sta}. Efficient scheduling requires timely knowledge of
buffered traffic, which the \gls{ap} can obtain implicitly from the QoS Control
field of transmitted frames or explicitly by polling \glspl{sta}. However,
implicit reporting cannot reveal newly buffered traffic at \glspl{sta} that have
not recently transmitted, while explicit polling is limited by the number of
available \glspl{ru}. Consequently, polling a \gls{sta} with an empty buffer may
delay the discovery of another with queued traffic. The polling policy therefore
determines how quickly the \gls{ap} can identify \glspl{sta} requiring \gls{ul}
resources~\cite{agbeve2025a2p}.

Several schemes have been proposed to address the buffer-status polling problem,
including \gls{sa-ofdma}, \gls{uora}, and
\gls{a2p}~\cite{agbeve2025evaluating,agbeve2025a2p}. These schemes have
generally been evaluated in isolation under different traffic and channel
conditions (cf.~\Cref{sec:related_works}). Consequently, their reported results
do not provide a controlled comparison of how the different access and polling
schemes perform across time-sensitive application scenarios with different
traffic patterns. In this work, we also introduce two alternative schemes,
\gls{a2s} and a \gls{mab}-based \gls{rl} polling scheme.

The main objective of this paper is to provide a controlled comparison of
\gls{edca}, \gls{sa-ofdma}, \gls{uora}, \gls{a2p}, \gls{a2s}, and \gls{rl}-based
polling using \mbox{ns-3} across three time-sensitive environments with
different traffic patterns and \gls{qos} requirements. These comprise (1) a
conferencing scenario, where sporadic talkers must regain service quickly, (2)
an \gls{iiot} deployment, where deadline-bound traffic coexists with persistent
background traffic, and (3) an \gls{xr} service, where the offered load is
dominated by \gls{dl} video while a smaller \gls{ul} head-tracking traffic has a
tighter deadline. Within each environment, all schemes are evaluated under the
same traffic and channel conditions, with their configurations tuned to achieve
the best performance among the evaluated settings.

Our main contributions are as follows.

\begin{itemize}

    \item We propose \gls{a2s}, an activity-based polling scheme that directly
        solicits \gls{ul} transmissions from recently active \glspl{sta} without
        first collecting their buffer-status reports.

    \item We formulate buffer-status polling as a \gls{cmab} problem, where each
        \gls{sta} is a base arm and the \gls{ap} selects a super-arm constrained
        by the number of available \glspl{ru}. We investigate multiple bandit
        algorithms and reward signals for learning which \glspl{sta} to poll
        from their observed buffer-status responses.

    \item We characterize the relative performance of contention-based,
        scheduled, random-access, activity-based, and learning-based access and
        polling schemes across three time-sensitive environments under common
        evaluation conditions. For each scheme and environment, we identify the
        best-performing configuration among the evaluated settings and provide
        the simulation code, configurations, and experiment scripts to enable
        reproduction of the reported
        results.\footnote{\url{https://github.com/imec-idlab/ns3-when-to-use-what}}

    \item Based on the observed performance limitations, we propose two
        multi-user scheduler enhancements tailored to the traffic asymmetries of
        the evaluated environments. Buffer-status- and access-category-aware
        selection for the \gls{iiot} deployment and backlog-directed \gls{dl}
        serving for the \gls{dl}-dominated \gls{xr} service.
    
\end{itemize}

The remainder of this paper is organized as follows. \Cref{sec:related_works}
reviews related work on \gls{ofdma} scheduling and polling.
\Cref{sec:background} describes the schemes under study. \Cref{sec:scheduler}
presents the proposed scheduler enhancements.
\Cref{sec:system_models,sec:sim_setup} define the traffic models and simulation
setup for the three environments. \Cref{sec:results} presents the results, and
\cref{sec:conclusion} concludes the paper.

\section{Related Work}
\label{sec:related_works}

Most prior work on Wi-Fi \gls{ofdma} focuses on scheduling and resource
allocation once the \gls{ap} has obtained the buffer state of the associated
\glspl{sta}. Joint scheduling and resource allocation have been formulated as
optimization problems which are solved using sub-optimal divide-and-conquer
recursive algorithms~\cite{wang2018scheduling}, while classical policies such
as MaxRate, Proportional Fair, and Shortest Remaining Processing Time have been
adapted to \gls{ul} \gls{ofdma}~\cite{bankov2018ofdma}. Deadline-aware
scheduling has been studied using queueing information derived from reported
buffer states~\cite{inamullah2020will}, and access-category buffer occupancy has
been used to prioritize resources for time-sensitive
traffic~\cite{qadri2022preparing}. Learning-based approaches have subsequently
been applied to \gls{ofdma} scheduling and joint \gls{ofdma}/MU-MIMO resource
allocation~\cite{noh2024joint,tan2024deep,quadri2025marl}. For \gls{dl} traffic,
latency-driven dynamic \gls{ofdma} scheduling has also been demonstrated on
commercial Wi-Fi~6 hardware~\cite{tai2025lomas}. These approaches improve
resource allocation when traffic demand or channel state is already available to
the scheduler, but acquiring recent buffer information remains a separate
problem for \gls{ul} scheduling.

One method for discovering \gls{ul} traffic is to use the \gls{ra}
\glspl{ru} provided by \gls{uora}. Early studies characterize its operation and
performance, and examine how successful \gls{ra} transmissions can provide
buffer information for subsequent scheduled
allocation~\cite{naik2018performace,bhattarai2019uplink}. Several extensions
modify the \gls{obo} selection or contention
window~\cite{rehman2023collision,kim2021ofdma,liu2025performance}, refine the
backoff
countdown~\cite{kosek2022efficient,kosek2022improving,rehman2025enhancing},
adapt the division between \gls{sa} and \gls{ra}
\glspl{ru}~\cite{bhattarai2019uplink,avdotin2020resource}, or introduce explicit
collision-resolution
mechanisms~\cite{avdotin2019ofdma,avdotin2019enabling,xie2020multi}. More recent
work considers \gls{uora} under freshness-sensitive traffic by optimizing its
parameters with respect to Age of Information~\cite{liu2025aoiuora}. Other
proposals reuse residual \gls{ul} transmission time after early completion or
collision~\cite{lee2025temporal}, control \gls{ru} contention and utilization
under variable \gls{iot} traffic~\cite{ouldamara2026prucc}, or make \gls{ra}
\gls{ru} selection dependent on traffic conditions~\cite{kim2026trafficaware}.
\Gls{uora} has also been combined with multi-link
operation~\cite{jin2024enhancing} and target wake
time~\cite{chen2020scheduling} to improve the performance. These works primarily
improve the operation of \gls{uora}. The complementary polling problem is how
quickly the \gls{ap} can discover that a \gls{sta} has new buffered traffic and
requires \gls{ul} resources.

Scheduled and hybrid schemes instead reduce dependence on \gls{ra} for
discovering and serving buffered traffic. Deterministic channel access can
centrally coordinate \gls{ul} transmissions while disabling random
access~\cite{schneider2023deterministic}, whereas multiple \gls{bsrp} exchanges
can collect additional buffer information before allocating transmission
resources~\cite{shao2024access}. Hybrid approaches allow a \gls{sta} to move
between contention-based and \gls{sa} according to its buffer
state~\cite{goncalves2023access}. Our earlier \gls{a2p} scheme follows this
approach by combining \gls{edca} and \gls{sa} \gls{ofdma} through a polling
list of recently active \glspl{sta}. Listed \glspl{sta} are subsequently served
through scheduled \gls{ul} \gls{ofdma} while their contention-based access is
suppressed through \gls{mu-edca} parameters~\cite{agbeve2025a2p}. Recent
experimental work further shows that trigger-based MU-\gls{ofdma} can provide
more predictable communication for industrial traffic when \gls{sa}, and timing
mechanisms are jointly optimized on commercial
hardware~\cite{orozcosantos2026deterministic}. Bandit-based methods have also
been applied to \gls{sta} selection. An \gls{mab}-based scheduler has been
proposed to prioritize \glspl{sta} according to information freshness, using Age
of Information as the scheduling objective~\cite{balci2025mabpolling}.

These research directions address different parts of the channel access and
scheduling process and are generally evaluated under different traffic, channel,
and load assumptions. Their results therefore cannot be directly compared to
determine how polling and channel access schemes behave across different
time-sensitive workloads. In earlier work, we evaluated \gls{uora} as a polling
scheme for latency-sensitive \gls{ul} traffic against \gls{sa-ofdma},
\gls{a2p}, and \gls{edca}, including the effect of the \gls{ocw} on delay and
scalability~\cite{agbeve2025evaluating}, using our \mbox{ns-3} implementation of
\gls{uora}~\cite{agbeve2025design,UORAns3}. That study considered a single
traffic environment.

This work extends the comparison to three time-sensitive environments with
different traffic generation patterns: conferencing environment with sporadic
talkers, an \gls{iiot} deployment combining background and deadline-bound
traffic, and a \gls{dl}-dominated \gls{xr} service. Within each scenario, we
evaluate contention-based (\gls{edca}), scheduled (\gls{sa-ofdma}),
random-access (\gls{uora}), activity-based (\gls{a2p} and \gls{a2s}), and
bandit-based (\gls{cmab}) schemes under the same traffic and channel
conditions.


\section{Background: Polling Schemes in Wi-Fi}
\label{sec:background}

This section first describes \gls{ul} \gls{ofdma} operation and the
\gls{mu-edca} parameter set. It then presents the random-access, activity-based,
and \gls{rl}-based polling schemes considered in this work.

\subsection{SA UL OFDMA}
\label{subsec:saofdma}

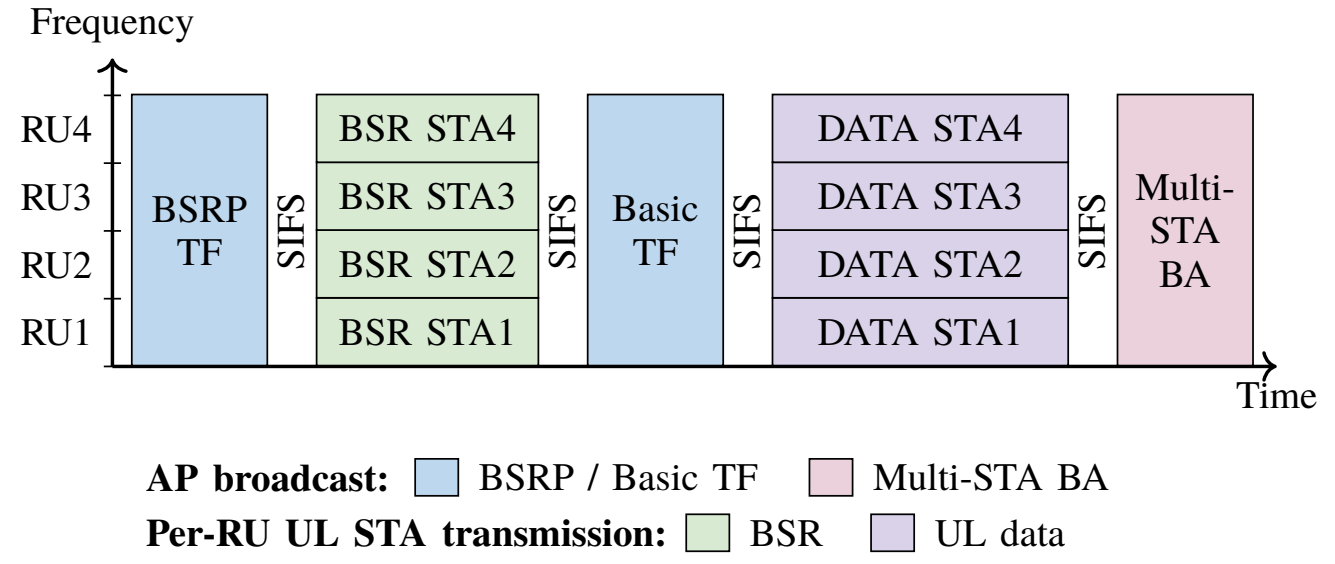
\begin{figure}[!t]
    \centering
    \tikzset{
        bcast/.style={rectangle, draw, thin, align=center,
                      fill=apBcast,   font=\small},
        bsrcell/.style={rectangle, draw, thin, align=center,
                        fill=phaseBSR, font=\small},
        ulcell/.style={rectangle, draw, thin, align=center,
                       fill=phaseUL,   font=\small},
        ackblk/.style={rectangle, draw, thin, align=center,
                       fill=phaseACK, font=\small},
        rulabel/.style={anchor=east, font=\small},
        sifslab/.style={rotate=90, anchor=center, font=\small},
    }
    \scalebox{0.815}{
    \begin{tikzpicture}[x=1cm, y=1cm]
        \def\hRU{0.55} \def\Htot{2.20}      
        \def\xA{0.00} \def\xB{1.10}         
        \def\xC{1.50} \def\xD{3.30}         
        \def\xE{3.70} \def\xF{4.80}         
        \def\xG{5.20} \def\xH{7.60}         
        \def\xI{8.00} \def\xJ{9.10}         

        \draw[->, thick] (-0.15,0) -- (-0.15,\Htot+0.3)
            node[above, font=\small]{Frequency};
        \draw[->, thick] (-0.15,0) -- (\xJ+0.20,0)
            node[below, font=\small]{Time};

        \foreach \i in {0,1,...,4}
            \draw[thin] (-0.22, {\i*\hRU}) -- (-0.08, {\i*\hRU});

        \foreach \i in {1,...,4}
            \node[rulabel] at (-0.19, {(\i-0.5)*\hRU}) {RU\i};

        \node[bcast, minimum width={(\xB-\xA)*1cm},
              minimum height={\Htot*1cm}, anchor=south west]
            at (\xA,0) {BSRP\\TF};

        \foreach \sta in {1,...,4}
            \node[bsrcell, minimum width={(\xD-\xC)*1cm},
                  minimum height=\hRU cm, anchor=south west]
                at (\xC,{(\sta-1)*\hRU}) {BSR STA\sta};

        \node[bcast, minimum width={(\xF-\xE)*1cm},
              minimum height={\Htot*1cm}, anchor=south west]
            at (\xE,0) {Basic\\TF};

        \foreach \sta in {1,...,4}
            \node[ulcell, minimum width={(\xH-\xG)*1cm},
                  minimum height=\hRU cm, anchor=south west]
                at (\xG,{(\sta-1)*\hRU}) {DATA STA\sta};

        \node[ackblk, minimum width={(\xJ-\xI)*1cm},
              minimum height={\Htot*1cm}, anchor=south west]
            at (\xI,0) {Multi-\\STA \\BA};

        \foreach \xl/\xr in {\xB/\xC, \xD/\xE, \xF/\xG, \xH/\xI}
            \node[sifslab] at ({(\xl+\xr)/2},{\Htot/2}) {SIFS};

        \def\yLegA{-0.9}     
        \def\yLegB{-1.35}    

        \node[anchor=west, font=\small\bfseries]
            at (0,\yLegA) {AP broadcast:};
        \node[bcast,  minimum width=.35cm, minimum height=.3cm,
              anchor=west] (L1) at (2.3,\yLegA) {};
        \node[right=1pt of L1, font=\small] {BSRP / Basic TF};
        \node[ackblk, minimum width=.35cm, minimum height=.3cm,
              anchor=west] (L2) at (5.5,\yLegA) {};
        \node[right=1pt of L2, font=\small] {Multi-STA BA};

        \node[anchor=west, font=\small\bfseries]
            at (0,\yLegB) {Per-RU UL STA transmission:};
        \node[bsrcell, minimum width=.35cm, minimum height=.3cm,
              anchor=west] (L3) at (4.5,\yLegB) {};
        \node[right=1pt of L3, font=\small] {BSR};
        \node[ulcell,  minimum width=.35cm, minimum height=.3cm,
              anchor=west] (L4) at (6.0,\yLegB) {};
        \node[right=1pt of L4, font=\small] {UL data};
    \end{tikzpicture}}
    \caption{UL OFDMA frame exchange sequence. Broadcast frames (BSRP TF, Basic
    TF, Multi-STA BA) occupy the full bandwidth; BSR and DATA transmissions
occur simultaneously on separate RUs.}
    \label{fig:ul_ofdma}
    \vspace{-.4cm}
\end{figure}
Under \gls{ofdma} operation, the \gls{ap} centrally coordinates channel access
by scheduling multiple \glspl{sta} and assigning each \gls{sta} a portion of the
available bandwidth, called an \gls{ru}. This bandwidth partitioning allows
multiple \glspl{sta} to communicate concurrently on separate \glspl{ru}. Wi-Fi
supports both \gls{dl} and \gls{ul} \gls{ofdma} transmissions. To orchestrate
either \gls{dl} or \gls{ul} \gls{ofdma} transmissions, the \gls{ap} first
acquires the channel through \gls{edca} contention.

During \gls{dl}, the \gls{ap} transmits concurrently to multiple \glspl{sta} over
separate \glspl{ru}. The \gls{ap} determines the \gls{ru} allocation according
to the queued \gls{dl} traffic and signals the mapping between \glspl{sta} and
\glspl{ru} as part of the multi-user transmission. The addressed \glspl{sta}
receive their respective data on the assigned \glspl{ru} and subsequently
acknowledge successful reception using the configured multi-user acknowledgment
procedure.

For \gls{ul} \gls{ofdma}, the \gls{ap} broadcasts a Basic \gls{tf} that specifies
\gls{ru} allocations for the selected \glspl{sta}. Following a \gls{sifs}, the
scheduled \glspl{sta} transmit simultaneously over their assigned \glspl{ru}.
Upon completion of the transmission, the \gls{ap} acknowledges the successfully
received packetswith a Multi-STA \gls{ba} after another \gls{sifs}. Efficient
\gls{ul} scheduling requires the \gls{ap} to know which \glspl{sta} have
buffered data. The \gls{ap} can acquire this information by sending a \gls{bsrp}
\gls{tf}, which contains, among other parameters, mappings of \glspl{sta} to
their assigned \glspl{ru}. The selected \glspl{sta}, identified by their
\glspl{aid} in the \gls{bsrp} \gls{tf}, respond with QoS Null frames carrying
their buffer-status information. A \gls{sta} can also report its buffer status
implicitly by including the buffer size in the QoS Control field of a
transmitted frame. Based on these reports, the \gls{ap} identifies the
\glspl{sta} with pending data and allocates resources to them in subsequent
\glspl{tf}. \Cref{fig:ul_ofdma} depicts the sequence of frame exchanges in
\gls{sa} \gls{ul} \gls{ofdma}.

To mitigate excessive contention in dense deployments, the standard introduces
the \gls{mu-edca} parameter set, which allows the \gls{ap} to regulate channel
access behavior of \glspl{sta} after their participation in \gls{ul} \gls{ofdma}.
The \gls{mu-edca} parameter set extends conventional \gls{edca} by enabling the
\gls{ap} to dynamically adjust contention parameters, such as the contention
window and the \gls{aifsn}, which are communicated via management frames. These
parameters are applied by \glspl{sta} during subsequent \gls{edca}-based channel
access. The parameter set also includes a timer that specifies the duration for
which the modified \gls{edca} behavior remains active. This timer is started or
refreshed when a \gls{sta} successfully completes an \gls{ul} \gls{ofdma}
transmission and receives a corresponding acknowledgement. If no further
successful transmissions occur before the timer expires, the \gls{sta} reverts
to its default \gls{edca} configuration.

\subsection{The UORA operation}

\tikzset{
    uora_sta/.style={rectangle, draw, thin, align=flush center,
                     fill=stasenddata,
                     minimum height=.5cm, minimum width=2.3cm, outer sep=0pt},
    uora_ap/.style={rectangle, draw, thin, align=flush center,
                    fill=mybgcolor,
                    minimum height=.5cm, minimum width=1cm},
    uora_ack/.style={rectangle, draw, thin, align=flush center,
                     fill=mybgcolor,
                     minimum height=2.5cm, minimum width=1cm},
    uora_collision/.style={fill=red!25},
    highlight_obo/.style={red, thick, rounded corners=1pt},
}

\gls{uora} enables contention-based access to a subset of \glspl{ru}, allowing
\glspl{sta} without explicit scheduled allocation to contend for transmission
opportunities. The AP initiates the \gls{ul} transmission by broadcasting a
\gls{tf}, which includes, among other information, the \glspl{ru} designated for
\gls{ra}. During association, the \gls{ap} advertises the \gls{ocw} range
through EOCW$_{\min}$ and EOCW$_{\max}$, which can subsequently be updated
through management frames. Each \gls{sta} derives its contention window
parameter bounds as $\mathrm{OCW}_{\min} = 2^{\mathrm{EOCW}_{\min}} - 1$ and
$\mathrm{OCW}_{\max} = 2^{\mathrm{EOCW}_{\max}} - 1$. The \gls{sta} initializes
its \gls{ocw} to $\mathrm{OCW}_{\min}$ and selects an \glsxtrfull{obo} value randomly
from the range $[0,\mathrm{OCW}]$. A \gls{sta} that is not explicitly assigned
an \gls{ru} but detects \gls{ra} opportunities in a received \gls{tf} decreases
its \gls{obo} by the number of available \gls{ra} \glspl{ru}. When the \gls{obo}
reaches zero or below, the \gls{sta} attempts transmission by randomly selecting
one of the available \gls{ra} \glspl{ru}. Following a successful transmission,
the \gls{ocw} is reset to $\mathrm{OCW}_{\min}$. After an unsuccessful
transmission, the \gls{sta} doubles its \gls{ocw}, up to $\mathrm{OCW}_{\max}$,
and draws a new \gls{obo} value from the updated contention window.

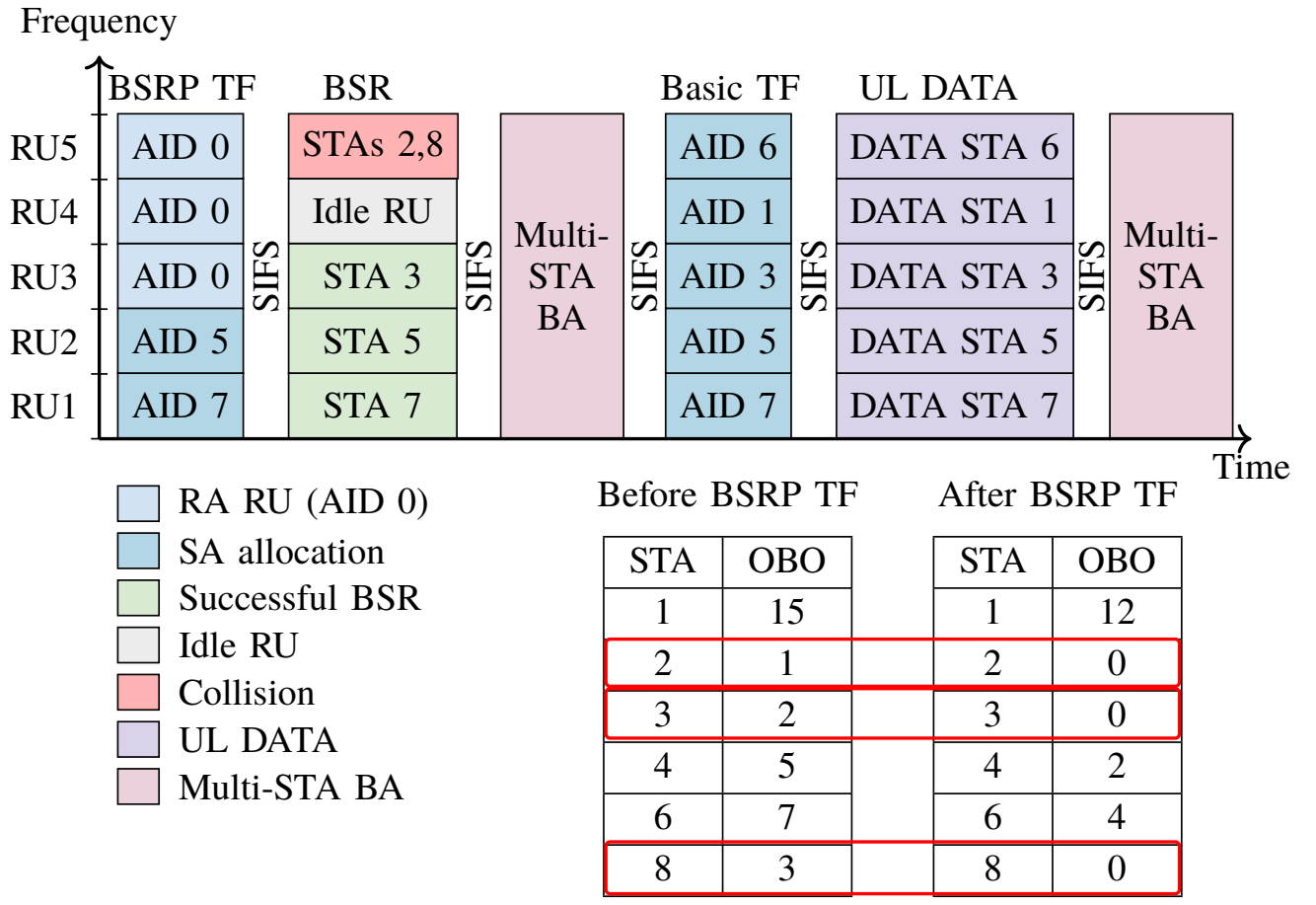
\begin{figure}[!t]
    \centering
    \tikzset{
        sacell/.style={rectangle, draw, thin, align=center,
                       fill=mybgcolor, font=\small},
        racell/.style={rectangle, draw, thin, align=center,
                       fill=apBcast!70, font=\small},
        bsrcell/.style={rectangle, draw, thin, align=center,
                        fill=phaseBSR, font=\small},
        idlecell/.style={rectangle, draw, thin, align=center,
                         fill=gray!15, font=\small},
        colcell/.style={rectangle, draw, thin, align=center,
                        fill=red!30, font=\small},
        ulcell/.style={rectangle, draw, thin, align=center,
                       fill=phaseUL, font=\small},
        ackblk/.style={rectangle, draw, thin, align=center,
                       fill=phaseACK, font=\small},
        rulabel/.style={anchor=east, font=\small},
        sifslab/.style={rotate=90, anchor=center, font=\small},
    }
    \scalebox{0.79}{
    \begin{tikzpicture}[x=1cm, y=1cm]
        \def\hRU{0.55}
        \def\Htot{2.75}   
        \def\xA{0.00} \def\xB{1.10}   
        \def\xC{1.45} \def\xD{2.64}   
        \def\xE{3.25} \def\xF{4.00}   
        \def\xG{4.65} \def\xH{5.50}   
        \def\xI{6.10} \def\xJ{7.85}   
        \def\xK{8.42} \def\xL{9.05}   

        \draw[->, thick] (-0.15,0) -- (-0.15,\Htot+0.5)
            node[above, font=\small]{Frequency};
        \draw[->, thick] (-0.15,0) -- (\xL+0.58,0)
            node[below, font=\small]{Time};

        \foreach \i in {0,1,...,5}
            \draw[thin] (-0.22,{\i*\hRU}) -- (-0.08,{\i*\hRU});
        \foreach \i in {1,...,5}
            \node[rulabel] at (-0.2,{(\i-0.5)*\hRU}) {RU\i};

        \foreach \row/\sty/\lbl in {%
            0/sacell/AID 7,
            1/sacell/AID 5,
            2/racell/AID 0,
            3/racell/AID 0,
            4/racell/AID 0} {
            \node[\sty, minimum width={(\xB-\xA)*0.5cm},
                  minimum height=\hRU cm, anchor=south west]
                at (\xA,{\row*\hRU}) (bsrp\row) {\lbl};
        }
        \node[above, font=\small] at ({(\xA+\xB)/2},\Htot) {BSRP TF};

        \node[bsrcell, minimum width={(\xD-\xC)*1.2cm},
              minimum height=\hRU cm, anchor=south west]
            at (\xC,{0*\hRU}) (bsr0) {STA 7};
        \node[bsrcell, minimum width={(\xD-\xC)*1.2cm},
              minimum height=\hRU cm, anchor=south west]
            at (\xC,{1*\hRU}) (bsr1) {STA 5};
        \node[bsrcell, minimum width={(\xD-\xC)*1.2cm},
              minimum height=\hRU cm, anchor=south west]
            at (\xC,{2*\hRU}) (bsr2) {STA 3};
        \node[idlecell, minimum width={(\xD-\xC)*1.2cm},
              minimum height=\hRU cm, anchor=south west]
            at (\xC,{3*\hRU}) (bsr3) {Idle RU};
        \node[colcell, minimum width={(\xD-\xC)*1.2cm},
              minimum height=\hRU cm, anchor=south west]
            at (\xC,{4*\hRU}) (bsr4) {STAs 2,8};
        \node[above, font=\small] at ({(\xC+\xD)/2},\Htot) {BSR};

        \node[ackblk, minimum width={(\xF-\xE)*1cm},
              minimum height={\Htot*1cm}, anchor=south west]
            at (\xE,0) (BAone) {Multi-\\STA \\ BA};

        \foreach \row/\lbl in {0/AID 7, 1/AID 5, 2/AID 3, 3/AID 1, 4/AID 6} {
            \node[sacell, minimum width={(\xH-\xG)*0.5cm},
                  minimum height=\hRU cm, anchor=south west]
                at (\xG,{\row*\hRU}) (btf\row) {\lbl};
        }
        \node[above, font=\small] at ({(\xG+\xH)/1.95},\Htot) {Basic TF};

        \foreach \row/\lbl in {%
            0/DATA STA 7, 1/DATA STA 5, 2/DATA STA 3,
            3/DATA STA 1, 4/DATA STA 6} {
            \node[ulcell, minimum width={(\xJ-\xI)*1cm},
                  minimum height=\hRU cm, anchor=south west]
                at (\xI,{\row*\hRU}) (data\row) {\lbl};
        }
        \node[above, font=\small] at ({(\xI+\xJ)/2},\Htot) {UL DATA};

        \node[ackblk, minimum width={(\xL-\xK)*1cm},
              minimum height={\Htot*1cm}, anchor=south west]
            at (\xK,0) (BAtwo) {Multi-\\STA\\ BA};

        \node[sifslab] at ($(bsrp2.east)!0.5!(bsr2.west)$)   {SIFS};
        \node[sifslab] at ($(bsr2.east)!0.5!(BAone.west)$)   {SIFS};
        \node[sifslab] at ($(BAone.east)!0.5!(btf2.west)$)   {SIFS};
        \node[sifslab] at ($(btf2.east)!0.5!(data2.west)$)   {SIFS};
        \node[sifslab] at ($(data2.east)!0.5!(BAtwo.west)$)  {SIFS};

        \def\yBot{-0.4}        
        \node[racell,  minimum width=.35cm, minimum height=.3cm,
              anchor=west] (Lg1) at (0,{\yBot-0.15}) {};
        \node[right=1pt of Lg1, font=\small] {RA RU (AID 0)};
        \node[sacell,  minimum width=.35cm, minimum height=.3cm,
              anchor=west] (Lg2) at (0,{\yBot-0.55}) {};
        \node[right=1pt of Lg2, font=\small] {SA allocation};
        \node[bsrcell, minimum width=.35cm, minimum height=.3cm,
              anchor=west] (Lg3) at (0,{\yBot-0.95}) {};
        \node[right=1pt of Lg3, font=\small] {Successful BSR};
        \node[idlecell,minimum width=.35cm, minimum height=.3cm,
              anchor=west] (Lg4) at (0,{\yBot-1.35}) {};
        \node[right=1pt of Lg4, font=\small] {Idle RU};
        \node[colcell, minimum width=.35cm, minimum height=.3cm,
              anchor=west] (Lg5) at (0,{\yBot-1.75}) {};
        \node[right=1pt of Lg5, font=\small] {Collision};
        \node[ulcell,  minimum width=.35cm, minimum height=.3cm,
              anchor=west] (Lg6) at (0,{\yBot-2.15}) {};
        \node[right=1pt of Lg6, font=\small] {UL DATA};
        \node[ackblk,  minimum width=.35cm, minimum height=.3cm,
              anchor=west] (Lg7) at (0,{\yBot-2.55}) {};
        \node[right=1pt of Lg7, font=\small] {Multi-STA BA};

        \node[font=\small, anchor=north west]
            at (4.0,{\yBot-.3}) (obo_before) {%
            \renewcommand{\arraystretch}{1.2}%
            \begin{tabular}{|c|c|}
                \hline
                STA & OBO \\ \hline
                1 & 15 \\ \hline
                2 & 1  \\ \hline
                3 & 2  \\ \hline
                4 & 5  \\ \hline
                6 & 7  \\ \hline
                8 & 3  \\ \hline
            \end{tabular}%
        };
        \node[above, font=\small] at (obo_before.north) {Before BSRP TF};
        \node[font=\small, anchor=north west]
            at (6.8,{\yBot-0.3}) (obo_after) {%
            \renewcommand{\arraystretch}{1.2}%
            \begin{tabular}{|c|c|}
                \hline
                STA & OBO \\ \hline
                1 & 12 \\ \hline
                2 & 0  \\ \hline
                3 & 0  \\ \hline
                4 & 2  \\ \hline
                6 & 4  \\ \hline
                8 & 0  \\ \hline
            \end{tabular}%
        };
        \node[above, font=\small] at (obo_after.north) {After BSRP TF};

        \pgfmathsetmacro{\rh}{0.40}
        \draw[highlight_obo]
            ($(obo_before.north west)+(0.15,-2.5*\rh cm)$)
            rectangle
            ($(obo_after.north east)+(-0.15,-3.5*\rh cm)$);   
        \draw[highlight_obo]
            ($(obo_before.north west)+(0.15,-3.55*\rh cm)$)
            rectangle
            ($(obo_after.north east)+(-0.15,-4.6*\rh cm)$);   
        \draw[highlight_obo]
            ($(obo_before.north west)+(0.15,-6.8*\rh cm)$)
            rectangle
            ($(obo_after.north east)+(-0.15,-7.9*\rh cm)$);   
    \end{tikzpicture}}
    \caption{\gls{ul} \gls{ofdma} frame exchange with \gls{uora}. The BSRP TF
    announces per-RU allocations: AID 0 marks \gls{ra} opportunities, non-zero
AIDs mark \gls{sa} assignments. STAs 2 and 8 pick the same \gls{ra} \gls{ru} and
collide (red). The tables show each STA's \glsxtrshort{obo} before and after the BSRP
TF; rows whose \glsxtrshort{obo} reaches zero are framed in red.}
    \label{fig:uora_ofdma}
    \vspace{-.4cm}
\end{figure}

\Cref{fig:uora_ofdma} illustrates the \gls{uora} frame-exchange sequence. The
\gls{ap} initiates the exchange by transmitting a \gls{bsrp} \gls{tf} containing
two \gls{sa} \glspl{ru}, assigned to \glspl{sta}~5 and~7, and three \gls{ra}
\glspl{ru}, identified by \gls{aid}~0. The three \gls{ra} opportunities reduce the
\gls{obo} values of \glspl{sta}~2, 3, and~8 from \num{1}, \num{2}, and \num{3} to
zero, making them eligible to transmit. After a \gls{sifs}, \glspl{sta}~5 and~7
send their \glspl{bsr} on their assigned \gls{sa} \glspl{ru}, while
\glspl{sta}~2, 3, and~8 select among the three \gls{ra} \glspl{ru}. \gls{sta}~3
transmits successfully, whereas \glspl{sta}~2 and~8 select the same \gls{ra}
\gls{ru} and collide. The remaining \gls{ra} \gls{ru} is idle. The \gls{ap}
acknowledges the successfully received reports with a Multi-STA \gls{ba} after a
\gls{sifs}. The subsequent Basic \gls{tf} assigns \glspl{ru} to \glspl{sta}~1,
3, 5, 6, and~7. \glspl{sta}~3, 5, and~7 have successfully reported their buffer
status in the preceding exchange, while \glspl{sta}~1 and~6 are scheduled using
buffer-status information already available at the \gls{ap}. After a \gls{sifs},
the five scheduled \glspl{sta} transmit their \gls{ul} data simultaneously on
the assigned \glspl{ru}. The exchange concludes with a Multi-STA \gls{ba}
transmitted by the \gls{ap} after the final \gls{sifs}. For clarity, the figure
shows equal-sized \glspl{ru}, although their sizes may differ in operation.

\subsection{\gls{a2p} and \gls{a2s}}
\label{subsec:a2panda2pv2}

In our earlier work~\cite{agbeve2025a2p}, we proposed \gls{a2p}, which combines
\gls{edca} and \gls{ofdma} to improve resource utilization in networks with a
large number of mostly idle \glspl{sta} that generate sporadic \gls{ul} traffic.
The \gls{ap} maintains a polling list of \glspl{sta} expected to have \gls{ul}
traffic. A \gls{sta} enters the polling list after initiating a successful
transmission through \gls{edca}. Following this transmission, the \gls{sta}
applies the \gls{mu-edca} mechanism described earlier, using the parameters
advertised by the \gls{ap} to disable its \gls{edca} access for a preconfigured
interval. During this interval, the \gls{sta} is served through scheduled
\gls{ul} \gls{ofdma} transmissions. When polled, the \gls{sta} reports its
buffer status and transmits pending data according to the allocated resources.
If the \gls{sta} remains inactive for the configured interval, the \gls{ap}
removes it from the polling list, after which the \gls{sta} resumes access
through \gls{edca}.

To reduce the polling overhead of \gls{a2p}, we propose \gls{a2s}, which retains
the same polling-list and \gls{mu-edca} mechanisms but removes the explicit
buffer-status polling step. Instead, the \gls{ap} selects \glspl{sta} from the
polling list using round-robin scheduling or the scheduler enhancements
described in \Cref{sec:scheduler}, assigning each selected \gls{sta} an \gls{ru}
through a Basic \gls{tf} for \gls{ul} data transmission. The \glspl{sta} report
their buffer status implicitly through the QoS Control field of each transmitted
QoS data frame.

\subsection{The \gls{rl} approach}
\label{subsec:cmab}

We further introduce a learning-based buffer-status polling approach. We
formulate the selection of \glspl{sta} to poll as a stochastic \glsxtrfull{cmab}
problem~\cite{Chen2013CUCB}. Let $\mathcal{N}=\{1,\dots,N\}$ denote the set of
associated \glspl{sta}, with each \gls{sta} corresponding to a base arm. At
polling round $t \in \{1,\dots,T\}$, where $T$ is the total number of polling
rounds, the \gls{ap} selects a super-arm
$\mathcal{S}_t\subseteq\mathcal{N}$ with cardinality
\[
    |\mathcal{S}_t|
    =
    K_t
    =
    \min(N,R_t),
\]
where $R_t$ is the number of \glspl{ru} available in round $t$.

Each selected arm $i\in\mathcal{S}_t$ yields a reward $r_{i,t}$
computed from the buffer-size field contained in the QoS~Null response
to the \gls{bsrp} \gls{tf} (\Cref{subsec:saofdma}). The reward depends
on the reward function used. We consider two reward functions, binary
and fairness-weighted. The binary reward is defined as
\begin{equation}
    r_{i,t}^{\mathrm{bin}}
    =
    \begin{cases}
        1, & \text{if } B_{i,t}>0,\\
        0, & \text{otherwise},
    \end{cases}
\end{equation}
where $B_{i,t}$ is the buffer size reported by \gls{sta} $i$ at
round $t$. The fairness-weighted reward additionally considers the
number of times a \gls{sta} has previously been selected, penalising
frequently selected \glspl{sta} while accounting for their reported
buffer sizes. Its reward can be expressed as
\begin{equation}
    r_{i,t}^{\mathrm{fair}}
    =
    f(B_{i,t},n_{i,t}),
\end{equation}
where $n_{i,t}$ denotes the number of previous selections of
\gls{sta} $i$, and $f$ is the fairness-weighted reward function
defined in \cref{appen:cmab}.

Let $\mu_{i,t}=\mathbb{E}[r_{i,t}]$ denote the expected reward of
arm $i$ at round $t$, under the selected reward function. The
expected reward of a super-arm is defined as
\begin{equation}
    \mu_t(\mathcal{S})
    =
    \sum_{i\in\mathcal{S}}\mu_{i,t}.
\end{equation}
The optimal feasible super-arm at round $t$ is
\begin{equation}
    \mathcal{S}_t^\star
    =
    \operatorname*{arg\,max}_{\substack{\mathcal{S}\subseteq\mathcal{N}\\
    |\mathcal{S}|=K_t}}
    \mu_t(\mathcal{S}).
\end{equation}
The objective is to minimise the cumulative pseudo-regret
\begin{equation}
    \mathcal{R}_T
    =
    \mathbb{E}\!\left[
    \sum_{t=1}^{T}
    \left(
        \mu_t(\mathcal{S}_t^\star)
        -
        \mu_t(\mathcal{S}_t)
    \right)
    \right],
    \label{eq:regret_sec}
\end{equation}
which measures the expected reward lost relative to selecting the
best feasible super-arm at each polling round.

We evaluate four selection algorithms under this formulation.
\Gls{cucb} selects arms using upper confidence bounds that balance
the estimated rewards with uncertainty, while \gls{cts} uses
posterior sampling to guide arm selection. \Gls{exp3m} maintains
exponential weights to adapt arm selection based on observed
rewards, whereas $\varepsilon$-greedy selects arms with the
highest estimated rewards while occasionally exploring other
arms. The detailed algorithm and reward formulations are
provided in \cref{appen:cmab}.

\tikzset{
  sysmodel/.style={
    >={Latex[length=1.8mm]},
    font=\small,
  },
  icon/.style={font=\fontsize{20}{20}\selectfont, inner sep=1pt},
  wired/.style={double, thick},
  ul/.style={->, very thick},
  dl/.style={->, very thick, aqua},
  lbl/.style={font=\footnotesize, align=center},
}

\section{Multi-user Scheduler Enhancements}
\label{sec:scheduler}

Based on the performance limitations observed during the evaluation, we introduce
two enhancements to the \gls{ap}'s \gls{ofdma}-based multi-user scheduler in
ns-3, targeting the traffic characteristics of the \gls{iiot} and \gls{xr}
scenarios.s

For \gls{iiot}, the scheduler must rapidly detect newly queued, deadline-bound
\gls{ul} frames while continuing to serve the persistent background traffic. In
the original implementation, a credit-based list determines the selection order
for both \gls{bsrp} polling and Basic \gls{tf} resource allocation. After each
Basic \gls{tf} exchange, selected \glspl{sta} earn credits proportional to the
transmission duration, while transmitting \glspl{sta} pay credits according to
their resource usage. The list is then re-ordered in decreasing credit order,
moving transmitting \glspl{sta} toward the back. This approach has two
limitations under the \gls{iiot} traffic mix. Firstly, polling does not consider
cached buffer status, so resources may be spent polling \glspl{sta} with known
backlog while newly queued traffic at other \glspl{sta} remains undiscovered.
Secondly, Basic \gls{tf} resource allocation considers cached buffer status to
exclude known-empty \glspl{sta} but ignores traffic priority, potentially
scheduling deadline-bound traffic after persistent background traffic.

We therefore modify the selection order at both stages. During \gls{bsrp}
polling, the \gls{ap} orders candidate \glspl{sta} by ascending cached buffer
size, treating unknown states as zero. This gives lower polling priority to
\glspl{sta} with known buffer size, allowing polling resources to focus on
discovering newly buffered traffic. During the subsequent Basic \gls{tf}
\gls{ru} allocation, the \gls{ap} orders \glspl{sta} according to the
highest-priority \gls{ac} with a positive \gls{bsr}, following $\mathrm{VO} >
\mathrm{VI} > \mathrm{BE} > \mathrm{BK}$. Round-robin ordering breaks ties among
equally prioritized \glspl{sta}, allowing deadline-bound traffic to be
prioritized over background traffic while maintaining round-robin service among
\glspl{sta} with the same priority.

The offered load in the \gls{xr} scenario is dominated by \gls{dl} video
traffic, while deadline-bound \gls{ul} head-tracking traffic constitutes only a
small fraction of the load. The reference ns-3 scheduler alternates strictly
between one \gls{dl} access and one \gls{ul} trigger exchange. The \gls{ap} can
also request channel access periodically, at intervals defined by the \gls{ari},
using the configured \gls{ac}, voice in our case, regardless of its queue state.
Each \gls{dl} transmission is restricted to \glspl{ac} with a priority equal to
or higher than that of the \gls{ac} that obtained the \gls{txop}. There are two
limitations to current approach under the \gls{xr} traffic. Firstly, strict
alternation allocates every other access to the \gls{ul}, despite its relatively
low offered load, reducing the transmission opportunities available to video
traffic. Secondly, \gls{dl} accesses may go unused. When a periodic channel
access request on the voice \gls{ac} obtains a \gls{txop} while no voice frames
are queued, the scheduler cannot serve the buffered video frames belonging to a
lower-priority \gls{ac}. Consequently, the \gls{dl} \gls{txop} remains unused
even when video traffic is buffered.

To address these limitations, we modify the \gls{dl}--\gls{ul} scheduling in two
ways. Firstly, the scheduler serves buffered \gls{dl} traffic on each access
obtained by an \gls{ac} with queued frames, while reserving the periodic
\gls{ari} accesses for \gls{ul} trigger exchanges. The \gls{ul} head-tracking
traffic is therefore served at intervals determined by the \gls{ari} rather than
consuming every other \gls{txop}. Secondly, we introduce backlog-directed
\gls{dl} \gls{ac} selection. When the \gls{ac} that obtained the \gls{txop} has
no queued frames and no \gls{ul} traffic to solicit, the \gls{ap} checks the
\glspl{ac} in the established priority order and schedules \gls{dl} traffic from
the first \gls{ac} with buffered frames. This preserves the priority of voice
traffic while allowing otherwise unused transmission opportunities to serve
queued video frames.

\section{System models}
\label{sec:system_models}

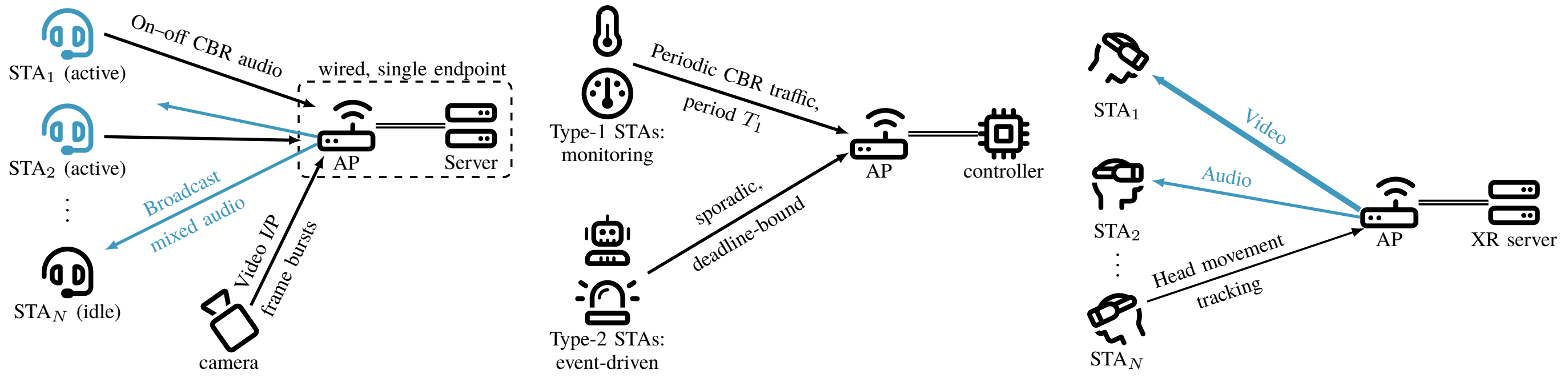
\begin{figure*}[!t]
  \centering
  \subfloat[Teleconferencing.\label{fig:sm_conf}]{%
  \scalebox{.85}{%
  \begin{tikzpicture}[sysmodel]
    \node[icon, text=aqua] at (0.8,2.5)  {\faIcon[regular]{headset}}; \node[lbl]
        at (0.8,1.95) {\gls{sta}$_1$ (active)};
    \node[icon, text=aqua] at (0.8,1.2)  {\faIcon[regular]{headset}}; \node[lbl]
        at (0.8,0.65) {\gls{sta}$_2$ (active)};
    \node at (0.8,0.2) {$\vdots$};
    \node[icon] at (0.8,-0.75) {\faIcon[regular]{headset}}; \node[lbl] at
        (0.8,-1.3)  {\gls{sta}$_{N}$ (idle)};
    \node[icon, xscale=-1, rotate=55] at (3.0,-1.4)  {\faIcon[regular]{video}};
    \node[lbl] at (3.0,-2.0)  {camera};
    \node[icon] at (4.6,1.25)  {\faIcon[regular]{router}}; \node[lbl] at
        (4.6,0.75)   {\gls{ap}};
    \node[icon] at (6.3,1.25)  {\faIcon[regular]{server}};  \node[lbl] at
        (6.3,0.75)   {Server};
    \draw[wired] (5.0,1.25) -- (5.95,1.25);
    \draw[dashed, thick, rounded corners=3pt] (3.95,0.55) rectangle (6.85,1.85);
    \node[lbl] at (5.5,2.0) {wired, single endpoint};
    \draw[ul] (1.3,2.5) -- (4.2,1.5)
      node[lbl, above=2pt, pos=0.45, sloped] {On--off \glsxtrshort{cbr} audio};
    \draw[ul] (1.3,1.1) -- (4.0,1.05);
    \draw[dl] (4.2,1.1) -- (2.0,1.55);
    \draw[dl] (4.2,1.0) -- (1.3,-0.45)
        node[lbl, above=0pt, pos=0.6, sloped] {Broadcast}
        node[lbl, below=0pt, pos=0.6, sloped] {mixed audio};
    \draw[ul] (3.3,-1.2) -- (4.3,0.85)
        node[lbl, above=0pt, pos=0.3, sloped] {Video I/P}
        node[lbl, below=0pt, pos=0.3, sloped] {frame bursts};
  \end{tikzpicture}}}
  \hfill
  \subfloat[Industrial \gls{iot}.\label{fig:sm_iiot}]{%
  \scalebox{.85}{%
  \begin{tikzpicture}[sysmodel]
    \node[icon] at (0.9,2.5)  {\faIcon[regular]{temperature-half}};
    \node[icon] at (0.9,1.65) {\faIcon[regular]{gauge}};
    \node[lbl]  at (0.9,0.95)  {Type-1 \glspl{sta}:\\ monitoring};
    \node[icon] at (0.9,-0.35) {\faIcon[regular]{user-robot}};
    \node[icon] at (0.9,-1.2)  {\faIcon[regular]{light-emergency-on}};
    \node[lbl]  at (0.9,-1.85) {Type-2 \glspl{sta}:\\ event-driven};
    \node[icon] at (4.6,1.1) {\faIcon[regular]{router}};    \node[lbl] at
        (4.6,0.6) {\gls{ap}};
    \node[icon] at (6.3,1.1) {\faIcon[regular]{microchip}}; \node[lbl] at
        (6.3,0.6) {controller};
    \draw[wired] (5.0,1.1) -- (5.95,1.1);
    \draw[ul] (1.25,2.05) -- (4.2,1.1)
      node[lbl, above=0pt, pos=0.45, sloped] {Periodic \glsxtrshort{cbr}
      traffic,}
     node[lbl, below=0pt, pos=0.45, sloped] {period $T_1$};
    \draw[ul] (1.45,-0.8) -- (4.2,0.85)
        node[lbl, above=0pt, pos=0.45, sloped] {sporadic,}
        node[lbl, below=0pt, pos=0.45, sloped] {deadline-bound};
  \end{tikzpicture}}}
  \hfill
  \subfloat[Extended reality.\label{fig:sm_xr}]{%
  \scalebox{.85}{%
  \begin{tikzpicture}[sysmodel]
      \node[icon, rotate=-35] at (0.9,3.3)
          {\faIcon[regular]{head-side-goggles}};
      \node[lbl] at (0.9,2.6) {\gls{sta}$_1$};

    \node[icon] at (0.9,1.6)  {\faIcon[regular]{head-side-goggles}};
    \node[lbl] at (0.9,0.95)  {\gls{sta}$_2$};

    \node at (0.9,0.6) {$\vdots$};
    \node[icon, xscale=-1, rotate=-25] at (0.9,-0.25)
        {\faIcon[regular]{head-side-goggles}}; \node[lbl] at (0.9,-0.8)
            {\gls{sta}$_N$};
    \node[icon] at (4.6,1.35) {\faIcon[regular]{router}}; \node[lbl] at
        (4.6,0.85) {\gls{ap}};
    \node[icon] at (6.3,1.35) {\faIcon[regular]{server}}; \node[lbl] at
        (6.3,0.85) {\gls{xr} server};
    \draw[wired] (5.0,1.35) -- (5.95,1.35);
    \draw[dl, line width=2.2pt] (4.2,1.25) -- (1.35,3.1) node[lbl, above=0pt,
        pos=0.5, sloped] {Video};
    \draw[dl] (4.2,1.15) -- (1.35,1.65) node[lbl, above=0pt, pos=0.65, sloped]
        {Audio};
    \draw[->, thick] (1.3,- 0.0) -- (4.27,1.0)
      node[lbl, above=0pt, pos=0.35, sloped] {Head movement}
      node[lbl, below=0pt, pos=0.35, sloped] {tracking};
  \end{tikzpicture}}}
  \caption{System models and traffic flows of the three environments. Arrows indicate traffic direction and their width indicates the share of the offered load.}
  \label{fig:system_models}
  \vspace{-.4cm}
\end{figure*}

We describe the system models of the three scenarios considered in this
study, namely a large-scale teleconferencing session, an industrial \gls{iot}
deployment, and an \gls{xr} service.

\subsection{Conferencing environment}

We model a teleconferencing session, illustrated in \cref{fig:sm_conf}, with
audio traffic following the model of our earlier work on
\gls{a2p}~\cite{agbeve2025a2p}. Participants communicate through a combined
microphone and headphone device. The microphone audio streams are transmitted
through an \gls{ap} to a server, which mixes the received \gls{ul} streams and
returns a combined \gls{dl} audio stream broadcast by the \gls{ap} to all
associated headphones. The \gls{ap}-to-server link is assumed to be
high-capacity and low-latency (e.g., fiber), so we treat the \gls{ap} and server
as a single logical endpoint. Each participant's microphone and headphones are
abstracted into one \gls{sta}. Although only a subset of \glspl{sta} is active
at any given time, the total number of associated \glspl{sta} can be large; for
example, parliamentary and debate scenarios may involve more than \num{100}
participants.

\begin{figure}[!t]
  \centering
  \scalebox{.75}{
  \begin{tikzpicture}[
      >={Latex[length=2mm]},
      font=\small,
      line/.style={very thick},
      pkt/.style={draw=aqua, very thick, fill=aqua!50, minimum height=5mm, minimum width=10mm},
      bcast/.style={draw, very thick, fill=gray!40, minimum height=5mm, minimum width=14mm},
      lost/.style={draw=red, very thick, fill=red!50, text=black, minimum height=5mm, minimum width=9mm},
      dashv/.style={dashed, thick}
      ]
        \def\xA{0}      
        \def\xB{1.0}    
        \def\xC{5.0}    
        \def\xD{6.55}    

        \def\yAP{2.5}
        \def\ySA{1.5}
        \def\ySB{0.75}
        \def\ySC{0.0}

        \foreach \y/\name in {\yAP/AP, \ySA/STA1, \ySB/STA2, \ySC/STA3}{
            \draw[->,line] (\xA-0.3,\y) -- (\xD+0.6,\y) node[right] {\name};
        }

        \foreach \x in {\xA,\xB,\xC,\xD}{
            \draw[dashv] (\x,-0.5) -- (\x,\yAP+0.85);
        }

        \draw[<->,thin] (\xA,\yAP-0.7) -- node[above]{B ms} (\xB,\yAP-0.7);

        \draw[<->,thin] (\xA,\yAP+0.5) -- node[above,pos=0.43]{Tx delay budget
            (X ms)} (\xC,\yAP+0.5);

        \draw[<->,thin] (\xB,\ySA+0.7) -- node[below, pos=.665]{Roundtrip delay}
            (\xD,\ySA+0.7);

        \node[bcast,anchor=south west] at (\xC,\yAP) {Broadcast};

        \node[pkt,anchor=south west] at (\xB+0.1,\ySA) {};

        \node[pkt,anchor= south west] at (\xB+1.8,\ySB) {};

        \node[lost,anchor=south] at (\xC+0.35,\ySC) {Lost};
        \draw[dashv] (\xC,\ySC) -- (\xC,\ySC+0.55);

        \draw[->,thick] (\xA+0.2,\ySA-0.5) -- (\xA+0.2,\ySA);   
        \draw[->,thick] (\xA+0.5,\ySB-0.5) -- (\xA+0.5,\ySB);   
        \draw[->,thick] (\xA+0.8,\ySC-0.7) -- (\xA+0.8,\ySC);       

        \node[below] at (\xB+0.5,\ySC-0.5) {Data generation time};

    \end{tikzpicture}
    }
    \caption{Traffic model of the teleconferencing scenario's audio stream.}
    \label{fig:system_model_b}
    \vspace{-.4cm}
\end{figure}
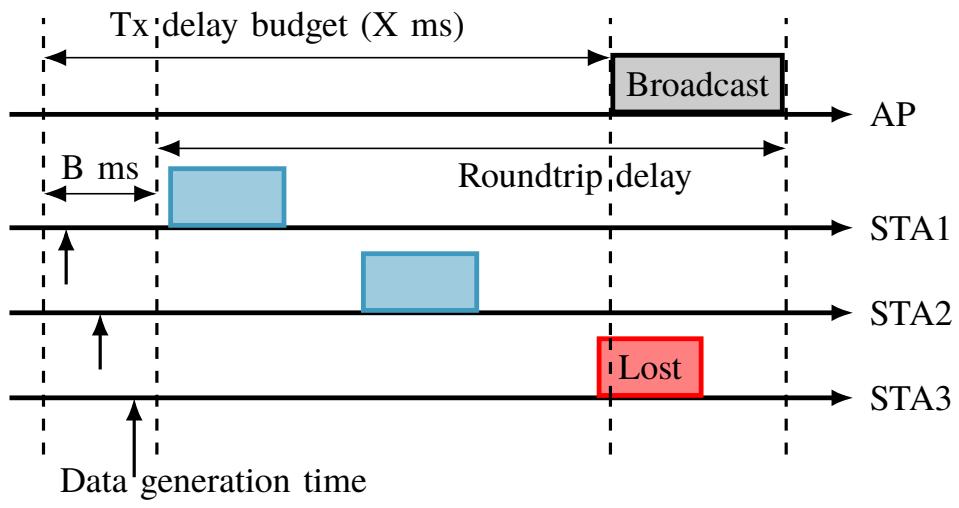

When a user activates the microphone, the \gls{sta} emits a \gls{cbr} \gls{ul}
audio stream that stops when the microphone is released, producing an on--off
traffic pattern. Time is divided into intervals of $X\,\unit{\milli\second}$.
Within the first $B\,\unit{\milli\second}$ of an interval, an \gls{sta} may
generate one audio packet of $M$ samples at $b$-bit resolution plus a $W$-bit
application header, giving a packet of $Mb + W\,\unit{\bit}$ and a rate of
$1000(Mb + W)/X\,\unit{\bps}$. Each packet must reach the \gls{ap} within a
delay budget of $X\,\unit{\milli\second}$, measured from generation to
reception; packets that arrive later are discarded by the server.
In~\cref{fig:system_model_b}, the packets from \gls{sta}~1 and \gls{sta}~2
arrive in time, whereas \gls{sta}~3's packet misses the budget and is dropped.

Once the server receives packets from all active \glspl{sta}, or the delay
budget expires, it forms a single \gls{dl} packet carrying the aggregated audio
of $P$ samples at $Q$-bit resolution plus a $W$-bit header, that is, $PQ +
W\,\unit{\bit}$ at rate $1000(PQ + W)/X\,\unit{\bps}$. This packet is broadcast
to every \gls{sta}. It carries only the audio that arrived in time (\gls{sta}~1
and \gls{sta}~2 in \cref{fig:system_model_b}), and its size does not depend on
the number of contributing stations.

One additional device is a video-streaming camera that projects the video of one
or more active speaker to a shared screen. The camera is wireless to allow it to
move and frame the relevant speaker or group of speakers in the hall. It is
therefore modelled as a \gls{sta} with an \gls{ul} video flow. In contrast, the
screen is fixed and connected to the server through the same high-speed wired
connection as the \gls{ap}, so it is treated as part of the \gls{ap}--server
endpoint. Consequently, only the camera's \gls{ul} video flow traverses the
wireless medium. Its traffic follows a low-latency H.264 encoder with frame rate
$F~\unit{\fps}$, where one I-frame is generated every $G$ frames and the
remaining $G-1$ frames are P-frames. Frame sizes are drawn independently from a
bounded Gaussian distribution. I-frame sizes are drawn from
$\mathcal{N}(\mu_I,\sigma_I^2)$ and bounded to $[\ell_I,u_I]$, while P-frame
sizes are drawn from $\mathcal{N}(\mu_P,\sigma_P^2)$ and bounded to
$[\ell_P,u_P]$~\cite{seeling2012video}. The corresponding mean frame sizes are
$\bar{B}_I$ and $\bar{B}_P$ bytes, resulting in mean offered video frame rate of
$ R = 8F\left[\frac{1}{G}\bar{B}_I+ \left(1-\frac{1}{G}\right)\bar{B}_P\right]
\,\unit{\bps},$ Each encoded frame is then packetized into \gls{udp} datagrams
carrying at most $S\,\unit{\bytes}$ of frame data, with datagrams belonging to
the same frame released at a fixed inter-datagram interval
$F_{t}~\unit{\second}$. The resulting traffic is bursty because the larger
I-frames recur every $G/F ~\unit{\second}$.

\subsection{Industrial environment}

We model a factory cell, depicted in \cref{fig:sm_iiot}, in which production
machinery and its instrumentation communicate wirelessly with an \gls{ap} that
forwards their data to an on-premise controller. The cell comprises two classes
of devices, referred to as \glspl{sta}, distinguished by their traffic
generation patterns. Deterministic (Type-1) \glspl{sta} represent
condition-monitoring sensors mounted on machinery, such as temperature probes,
vibration sensors, and position encoders, which report measurements
periodically. Stochastic (Type-2) \glspl{sta} represent event-driven devices,
such as safety sensors, fault detectors, and mobile robots reporting unexpected
obstructions. These devices generate sporadic reports that must reach the
controller within a specified deadline to enable timely action and are therefore
treated as the time-critical class.

Type-1 \glspl{sta} generate \gls{cbr} \gls{ul} \gls{udp} traffic with a period
$T_1$ and payload size $L~\unit{\byte}$, corresponding to an offered load of
$\frac{8L}{T_1}~\unit{\Mbps}$ per \gls{sta}. Type-2 \glspl{sta} generate
\gls{udp} packets of the same size, with exponentially distributed inter-arrival
times of mean $\mu$.

\subsection{\Gls{xr} environment}

The extended-reality (XR) model, illustrated in \cref{fig:sm_xr}, follows the
multi-stream traffic model of Carrascosa \emph{et~al.}~\cite{Carrascosa2024}.
Each headset is modeled as a single \gls{sta} carrying three concurrent flows.
On the \gls{dl}, the video flow releases a burst of $N_v$ datagrams of size
$S_v$ at the start of each video frame period $T_v$, while the audio flow
releases $N_a$ datagrams of size $S_a$ every audio period $T_a$. On the
\gls{ul}, the pose-tracking flow carries updates of the user's head position and
orientation to the \gls{xr} server. It generates $N_u$ datagrams during each
tracking period $T_u$, three of size $S_{u1}$ and one of size $S_{u2}$, so the
aggregate size per period is $S_u = 3S_{u1} + S_{u2}$. Unlike the \gls{dl}
flows, which release all their datagrams in a single burst at the start of each
period, the pose flow assigns each of its $N_u$ datagrams an independently
sampled release time, uniformly distributed over the tracking period. The
corresponding mean rates are $R_v = 8N_vS_v/T_v$, $R_a = 8N_aS_a/T_a$, and $R_u
= 8S_u/T_u$. The aggregate offered load is \gls{dl}-dominant, as the combined
video and audio rate exceeds the pose-tracking rate.

\section{Simulation setup}
\label{sec:sim_setup}

\begin{table}[!t]
    \centering
    \caption{Common \glsxtrshort{mac} and Scheme Configuration}
    \label{tab:sim_common}
    \begin{center}{
            \renewcommand{\arraystretch}{1.5}
            \begin{tabular}{l c}
                \hline
                \textbf{Parameter} & \textbf{Value} \\
                \hline
                Carrier frequency & \qty{5}{\giga\hertz} \\
                Guard Interval & \qty{0.8}{\micro\second} \\
                \glsxtrshort{sifs} / slot & \qty{16}{\micro\second} /
                \qty{9}{\micro\second} \\
                Propagation Loss Model & IEEE Indoor Model B \cite{indoorModel}
                \\
                \gls{txop} & \qty{2080}{\micro\second} \vspace{0.2cm}\\
                                \gls{mu-edca} timer &
                                \pbox{10cm}{\qty{40.96}{\milli\second}
                                    (\gls{a2p}, \gls{a2s}), \\ Sim. duration for
                                others \vspace{0.2cm}} \\
                \glsfmtshort{ocw}$_{\mathrm{min}}$ &
                                    \numlist[list-units=single,
                                    list-pair-separator={\,~}]{1;3;7;15;31;63}
                                    \\
                \glsfmtshort{ocw}$_{\mathrm{max}}$ & \num{127} \\
                 Duration of Simulation & \qty{30}{\second} \\
                \hline
        \end{tabular}}
    \end{center}
    \vspace{-.6cm}
\end{table}

We evaluate the polling schemes using the ns-3 network simulator~\cite{ns3}.
Each experiment models a single \gls{bss} comprising one \gls{ap} and multiple
\glspl{sta}, all compliant with the Wi-Fi 6 (or newer) standard. Each simulation
runs for \qty{30}{\second}. For \gls{a2p} and \gls{a2s}, we set the
\gls{mu-edca} timer to \qty{40.96}{\milli\second} and disable \gls{edca} for the
entirety of each experiment under the other \gls{ofdma}-based schemes. Under
\gls{uora}, we sweep the minimum \gls{ocw} over (\gls{ocw}$_{\mathrm{min}} =
\numlist[list-units=single, list-pair-separator={\,~}]{1;3;7;15;31;63}$) while
keeping the maximum \gls{ocw} at \num{127}. Since the \gls{ap} requires channel
access to coordinate \gls{ul} transmissions, it cannot rely solely on queued
\gls{dl} packets to initiate channel access. Instead, it requests access
periodically, at intervals defined by the \gls{ari}. For each scheme and
scenario, we sweep the \gls{ari} over power-of-two values between
\qty{16}{\micro\second} and \qty{1024}{\micro\second} and select the value that
minimizes the deadline loss of the deadline-bound traffic.In the presence of
overlapping \glspl{bss}, the selection of \gls{ari} would also need to account
for fairness in channel access across \glspl{bss}.

Additionally, we ensure that the transmission power is sufficiently high that
the probability of packet loss due to channel errors is negligibly small. Hence,
in the various setups the packets can be lost only due to collisions or not
delivered within their application deadline. The \gls{mac} and scheme parameters
common to all scenarios are listed in \Cref{tab:sim_common}. We next describe
the scenario-specific configurations. The ns-3 scenario implementations and the
scripts that generate all figures in this paper are publicly
available~\cite{pollingschemens3}.

\subsection{Conferencing environment}

\begin{table}[!t]
    \centering
    \caption{Conferencing Scenario Parameters}
    \label{tab:sim_conf}
    \renewcommand{\arraystretch}{1.5}
    \begin{tabular}{l c}
        \hline
        \textbf{Parameter} & \textbf{Value} \\
        \hline
        Bandwidth & \qty{40}{\mega\hertz} \\
        MCS  & 8 \\
        Resource Unit Type & 26\hyphen tone only \\
        UL Audio Generation Window Part, $B$ & \qty{1}{\milli\second} \\
        Off- to on- \& On- to off-time interval  & \pbox{10cm}{Bounded Exp.
                                    Distribution, \\(mean \qty{10}{\second},
                                    max \qty{25}{\second})} \vspace{0.2cm} \\
        \pbox{10cm}{Audio packet interval \\and delay budget}, $X$ &
                        \qty{5}{\milli\second} \vspace{0.2cm} \\
        Audio \gls{ul}/\gls{dl} Payload Size (AC\_VO) &
                                        \num{740}~/~\num{500}~\unit{\byte} \\
        AP \gls{ari}, $T$ & 
                            \pbox{3.5cm}{\vspace{0.2cm}\qty{16}{\micro\second}
                                (\gls{uora}, \gls{cmab}) \\
                                \qty{128}{\micro\second}(\gls{a2p}),
                                \qty{256}{\micro\second}(\gls{a2s})
                                        \vspace{0.1cm}}\\
        Video Stream (AC\_VI) & H.264, \num{30}~fps, GOP~\num{15} \\
            \qquad I / P frame, $[\ell, u]$ &
                    \pbox{10cm}{$\mathcal{N}(\num{80000},\num{12000}^2)$ / \\
                        $\mathcal{N}(\num{1500},\num{600}^2)$~\unit{\byte}} \\
            \qquad datagram payload, $S$ & \qty{1388}{\byte} \\
            \qquad inter-datagram interval, $F_t$ & \qty{50}{\micro\second}
                                                                            \\
            \qquad mean rate, $R$ & ${\approx}\,\qty{1.6}{\Mbps}$ \\
        Number of \glspl{sta} & \num{100} \\
        \hline
        \end{tabular}
    \vspace{-.4cm}
\end{table}

Each audio \gls{sta} generates \gls{cbr} \gls{udp} traffic (AC\_VO) at
approximately \qty{1.2}{\Mbps}. A \gls{ul} packet consists of $M=240$ audio
samples with $b=24$-bit resolution and includes a $W=20$-byte application-layer
header. The \gls{ap} broadcasts \gls{dl} \gls{udp} traffic to all associated
\glspl{sta} at approximately \qty{0.1}{\Mbps}, each \gls{dl} packet contains
$P=240$ audio samples with $Q=16$-bit resolution and the same $W=20$-byte
application-layer header. During each simulation run, the \num{100} \glspl{sta}
associated with the \gls{ap} are dynamically classified as either active, i.e.,
transmitting an audio stream, or idle. Before the start of the simulation,
\num{8} initial active \glspl{sta} and a variable number of joining \glspl{sta}
are randomly selected. The initial active \glspl{sta} generate their first
packets at uniformly random times within $B=\qty{1}{\milli\second}$ after system
initialization. Subsequently, each joining \gls{sta} independently draws its on-
and off-period durations from exponential distributions with a mean of
\qty{10}{\second} and an upper bound of \qty{25}{\second}.

The camera \gls{sta} generates \gls{ul} \gls{udp} video traffic (AC\_VI) at
$F=\num{30}$~fps with a group-of-pictures length of $G=\num{15}$ frames.
Following the model in \Cref{sec:system_models}, I-frame sizes use
$\mu_I=\qty{80}{\kilo\byte}$ and $\sigma_I=\qty{12}{\kilo\byte}$, while P-frame
sizes use $\mu_P=\qty{1500}{\byte}$ and $\sigma_P=\qty{600}{\byte}$. Frame sizes
are bounded to $[\ell_I{=}\num{500},u_I{=}\num{200000}]\,\unit{\byte}$ for
I-frames and
$[\ell_P{=}\num{500},u_P{=}\num{30000}]\,\unit{\byte}$ for P-frames. Each frame
is packetized into \gls{udp} datagrams carrying at most $S=\qty{1388}{\byte}$ of
frame data, with consecutive datagrams released \qty{50}{\micro\second} apart.
This configuration produces a mean video rate of \qty{1.62}{\Mbps}, with larger
I-frame bursts recurring twice per second. The conferencing scenario parameters
are listed in \Cref{tab:sim_conf}.

\subsection{Industrial environment}

\begin{table}[!t]
    \centering
    \caption{Industrial Scenario Parameters}
    \label{tab:sim_iiot}
    \renewcommand{\arraystretch}{1.5}
    \begin{tabular}{l c}
        \hline
        \textbf{Parameter} & \textbf{Value} \\
        \hline
        Bandwidth &  \qty{40}{\mega\hertz} \\
        MCS & \num{8} \\
        Resource Unit Type & 26\hyphen tone only \\
        EDCA Access Category & VO (\textit{Type-2}), BE (\textit{Type-1}) \\
        AP \gls{ari}, $T$ & \pbox{3.5cm}{\vspace{0.2cm}\qty{16}{\micro\second}
            (\gls{uora}, \gls{cmab}) \\ \qty{512}{\micro\second} (\gls{a2p}),
            \qty{256}{\micro\second} (\gls{a2s})
                                        \vspace{0.2cm}}\\
        Payload Size. $L$ & \qty{700}{\byte} \\
        Deterministic Inter-Packet Interval, $T_1$ & \qty{2.08}{\milli\second}
            \vspace{3pt} \\
                Stochastic Inter-Packet Interval  &
                \pbox{8cm}{Exp. Distribution, \\(means:
                \{\qtyrange[range-phrase={--},range-units=single]
                                        {12.5}{200}{\milli\second}\})} \\
        Type-2 delivery deadline ($D$) & \qty{20}{\milli\second} \\
        Number of Deterministic STAs & \num{18} \\
        \hline
    \end{tabular}
    \vspace{-.4cm}
\end{table}

For the \gls{iiot} evaluation, we configure \num{18} Type-1 \glspl{sta},
matching the \num{18} 26-tone \glspl{ru} available in the
\qty{40}{\mega\hertz} bandwidth. A single trigger frame can therefore serve
all Type-1 \glspl{sta}, whereas allocating an \gls{ru} to a Type-2 \gls{sta}
removes one scheduled transmission opportunity from the Type-1 traffic.
The resulting Type-1 offered load (AC\_BE) is approximately
\qty{48.5}{\Mbps}. We vary the Type-2 traffic load (AC\_VO) by sweeping the mean
inter-arrival time from \qty{12.5}{\milli\second} to \qty{200}{\milli\second}.
The remaining deployment and traffic parameters are summarized in
\Cref{tab:sim_iiot}.

\subsection{\Gls{xr} environment}

\begin{table}[!t]
    \centering
    \caption{Extended-Reality Scenario Parameters~\cite{Carrascosa2024}}
    \label{tab:sim_xr}
    \renewcommand{\arraystretch}{1.5}
    \begin{tabular}{l c}
        \hline
        \textbf{Parameter} & \textbf{Value} \\
        \hline
        Bandwidth & \qty{160}{\mega\hertz} \\
        MCS & 11 \\
        Resource Unit Type & 484\hyphen , 996\hyphen tone \\
        Number of Headsets (\gls{sta}) & \numrange{1}{8} \\
        AP \gls{ari}, $T$ & \pbox{3.5cm}{\vspace{0.2cm}\qty{256}{\micro\second}
            (\gls{uora}, \gls{cmab}) \\ \qty{128}{\micro\second} (\gls{a2p}),
            \qty{512}{\micro\second} (\gls{a2s})
                                        \vspace{0.2cm}}\\
        \multicolumn{2}{l}{DL Video (AC\_VI)} \\
        \qquad datagrams per frame, $N_v$ & \num{96} \\
        \qquad datagram size, $S_v$ & \qty{1448}{\byte} \\
        \qquad frame period, $T_v$ & \qty{11.11}{\milli\second} \\
        \qquad mean rate, $R_v$ & ${\approx}\,\qty{100}{\Mbps}$ \\

        \multicolumn{2}{l}{DL Audio (AC\_VO)} \\
        \qquad datagrams per period, $N_a$ & \num{4} \\
        \qquad datagram size, $S_a$ & \qty{1448}{\byte} \\
        \qquad audio period, $T_a$ & \qty{25}{\milli\second} \\
        \qquad mean rate, $R_a$ & ${\approx}\,\qty{1.85}{\Mbps}$ \\

        \multicolumn{2}{l}{UL Pose Estimate (AC\_VO)} \\
        \qquad datagrams per period, $N_u$ & \num{4} \\
        \qquad datagram sizes, $S_{u1}$ / $S_{u2}$ & \qty{106}{\byte} /
                                                \qty{212}{\byte} \\
        \qquad aggregate size, $S_u$ & \qty{530}{\byte} \\
        \qquad tracking period, $T_u$ & \qty{11.11}{\milli\second} \\
        \qquad mean rate, $R_u$ & ${\approx}\,\qty{0.38}{\Mbps}$ \\
        Video delivery deadline ($D_v$) & \qty{50}{\milli\second} \\
        Audio delivery deadline ($D_a$) & \qty{30}{\milli\second} \\
        Pose delivery deadline ($D_u$) & \qty{10}{\milli\second} \\
        \hline
    \end{tabular}
    \vspace{-.4cm}
\end{table}

A single \gls{ap} serves \numrange{1}{8} headsets (\glspl{sta}). A server
co-located with the \gls{ap} generates one \gls{dl} video flow and one \gls{dl}
audio flow per headset, while each headset generates its own \gls{ul}
pose-tracking flow. \Cref{tab:sim_xr} summarizes the traffic parameters. The
\gls{dl} video flow (AC\_VI) releases one encoded frame every
$T_v=\qty{11.11}{\milli\second}$, corresponding to \qty{90}{\fps}, as a burst of
$N_v=\num{96}$ datagrams of size $S_v=\qty{1448}{\byte}$. This yields
$R_v\approx\qty{100}{\Mbps}$ per headset. The \gls{dl} audio flow (AC\_VO)
releases $N_a=\num{4}$ datagrams of size $S_a=\qty{1448}{\byte}$ every
$T_a=\qty{25}{\milli\second}$, yielding $R_a\approx\qty{1.85}{\Mbps}$. The
\gls{ul} pose-tracking flow (AC\_VO) generates $N_u=\num{4}$ datagrams every
$T_u=\qty{11.11}{\milli\second}$, three of size $S_{u1}=\qty{106}{\byte}$ and
one of size $S_{u2}=\qty{212}{\byte}$. Their aggregate size is
$S_u=\qty{530}{\byte}$, yielding $R_u\approx\qty{0.38}{\Mbps}$.

Video and audio datagrams are released together at the start of their respective
periods and therefore enter the \gls{ap} queue as bursts. Pose datagrams are
instead released independently at uniformly distributed instants within each
tracking period. The resulting offered load is strongly \gls{dl}-dominated and
increases with the number of headsets. Given this high offered load, we use
\qty{160}{\mega\hertz} bandwidth and \gls{mcs}~\num{11}. Each \gls{ul} trigger
frame contains four 484-tone \glspl{ru}, with \numrange{0}{4} assigned to random
access. On the \gls{dl}, each headset is assigned a 996-tone \gls{ru} when at
most two headsets are served; from three headsets onward, 484-tone \glspl{ru}
are used instead. This allocation provides a larger bandwidth share per headset
at low headset counts. \Cref{tab:sim_xr} lists the \gls{xr} service parameters.

\section{Results and discussion}
\label{sec:results}

We compare the schemes using a common set of metrics namely Delay and the
fraction of generated packets that are either dropped from the queue or not
delivered in time. Let $\mathcal{P}$ denote the set of packets generated during
the measurement interval $T_\mathrm{m}$, and $\mathcal{R} \subseteq \mathcal{P}$
denote the subset that is delivered. The delay $d_p$ of a packet $p \in
\mathcal{R}$ is measured from its generation time at the source application to
its reception at the destination application. We set $d_p = \infty$, for an
undelivered packet $p \in \mathcal{P} \setminus \mathcal{R}$.

For deadline-bound traffic with deadline $D$, we use the packet deadline-miss
ratio
\begin{equation}
  \Lambda \;=\; 1 -
  \frac{\bigl|\{\, p \in \mathcal{P} : d_p \le D \,\}\bigr|}
       {|\mathcal{P}|}.
  \label{eq:dmr}
\end{equation}
A packet misses the deadline whether it arrives late or is never delivered, so
packet drops cannot falsely improve the metric. The deadline is
\qty{20}{\milli\second} for the industrial Type-2 flow~\cite{ieee80211rta} and
\qtylist{50;30;10}{\milli\second} for the \gls{xr} video, audio, and pose flows,
respectively ~\cite{wfa2023vr}. The corresponding deadline in the conferencing
scenario is the delay budget $X=\qty{5}{\milli\second}$. We consider a flow
served when its deadline-miss ratio does not exceed \num{0.05}.

For the conferencing scenario, we additionally report the wake-up delay of an
\gls{sta}, defined as the delay $d_{p^\ast}$ of the first packet $p^\ast$
generated after its microphone turns on. This metric measures how quickly a
scheme provides \gls{ul} access to a newly active \gls{sta}.

Delay tails are reported as percentiles over delivered packets. Let
$Q_q(\mathcal{A})$ denote the $q$-quantile of the delays
$\{ d_p : p \in \mathcal{A} \}$ of a packet set
$\mathcal{A} \subseteq \mathcal{R}$. For \gls{xr} we report the worst-station
\nthd{99.9}-percentile delay $\max_s Q_{0.999}(\mathcal{R}_s)$, where
$\mathcal{R}_s$ is the set of packets delivered for \gls{sta} $s$, so a
scheme cannot mask a poorly served headset behind an aggregate percentile.
For the industrial environment we report $Q_{0.99}(\mathcal{R})$ over all
Type-2 packets. Throughput is the application-layer rate
$\sum_{p \in \mathcal{R}} b_p \,/\, T_\mathrm{m}$, where $b_p$ is the payload
size of packet $p$ in bits.

\subsection{Conferencing environment}
\subsubsection{\glsentryshort{cmab} reward and hyperparameter selection}
\label{sec:conf_cmab}

\begin{figure*}[!t]
\centering
\subfloat[Packet loss ratio.]{%
  \includegraphics[width=0.98\textwidth]{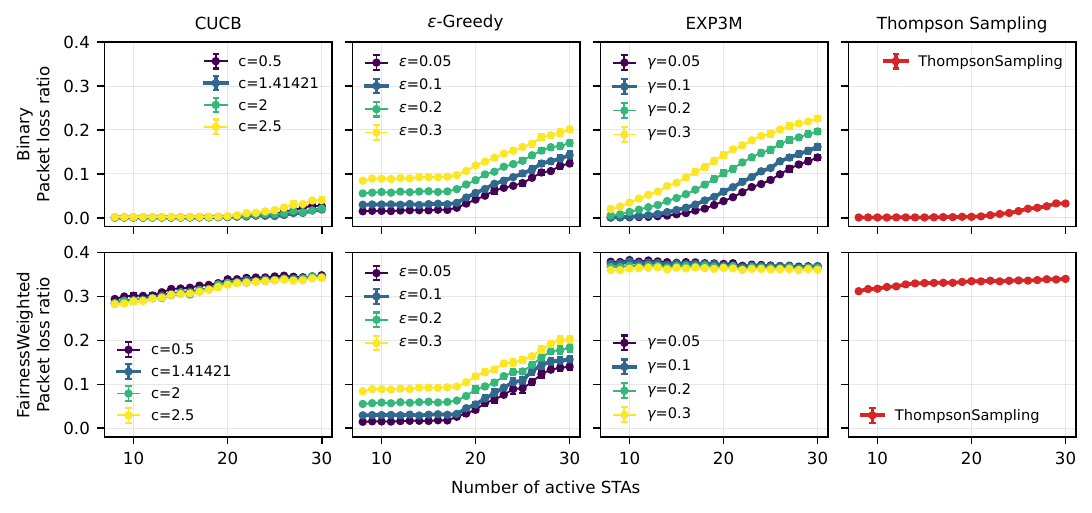}%
  \label{fig:cmab_loss}}\\
\subfloat[Average wake-up delay.]{%
  \includegraphics[width=0.98\textwidth]{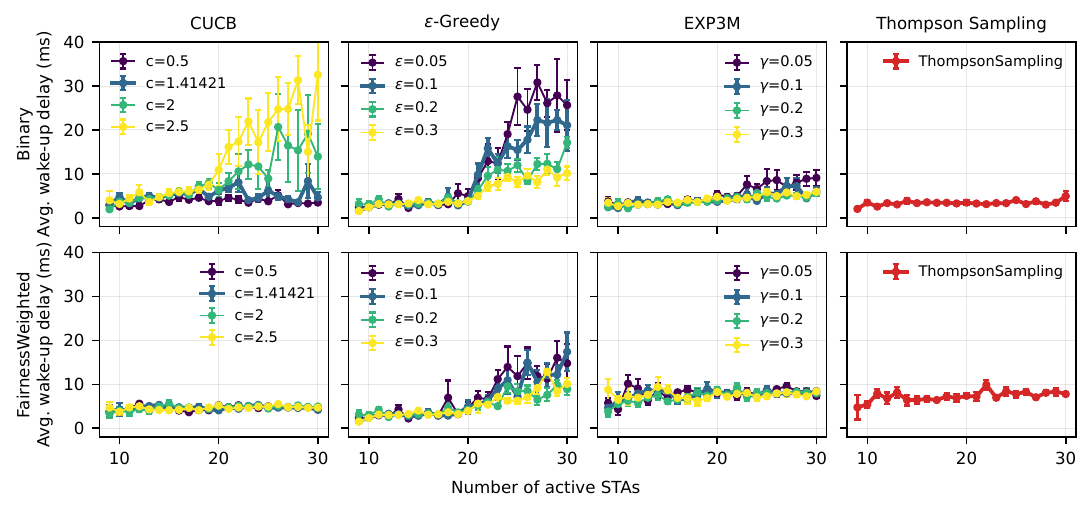}%
  \label{fig:cmab_wakeup}}
  \caption{\gls{cmab} hyperparameter sweep for the four algorithms (columns) as
  the number of active \glspl{sta} increases, evaluated under binary and
fairness-weighted reward functions (rows).}
\label{fig:cmab}
\end{figure*}

Before comparing \gls{cmab} with the baseline schemes, we select its reward
signal and algorithm-specific hyperparameter. \Cref{fig:cmab_loss} shows the
packet-loss ratio, while \cref{fig:cmab_wakeup} shows the mean wake-up delay for
the evaluated algorithms as the number of active \glspl{sta} increases. The
reward signal has the largest effect. The fairness-weighted reward spreads
\glspl{ru} across \glspl{sta} regardless of their cached buffer sizes and
therefore yields a considerably higher packet-loss ratio than the binary reward,
which credits every successful transmission and thereby favours \glspl{sta} with
queued traffic. Only $\varepsilon$-greedy is insensitive to this choice, showing
no difference in packet loss between the two rewards. Under the binary reward,
\gls{cucb} achieves the lowest packet-loss ratio among the four algorithms and
is the least sensitive to its exploration constant $c$, which trades packet loss
against wake-up delay. Reducing $c$ from $\sqrt{2}$ to \num{0.5} raises the
packet-loss ratio only from \num{0.019} to \num{0.027} at \num{30} active
stations while lowering the mean wake-up delay from \qty{4.6}{\milli\second} to
\qty{3.5}{\milli\second}. Consequently, we adopt \gls{cucb} with the binary
reward and $c=0.5$ for the remainder of the evaluation, as it achieves
near-minimum packet loss while minimizing wake-up delay.

\subsubsection{Scheme comparison}
\begin{figure*}[!t]
  \centering
  \includegraphics[width=\textwidth]{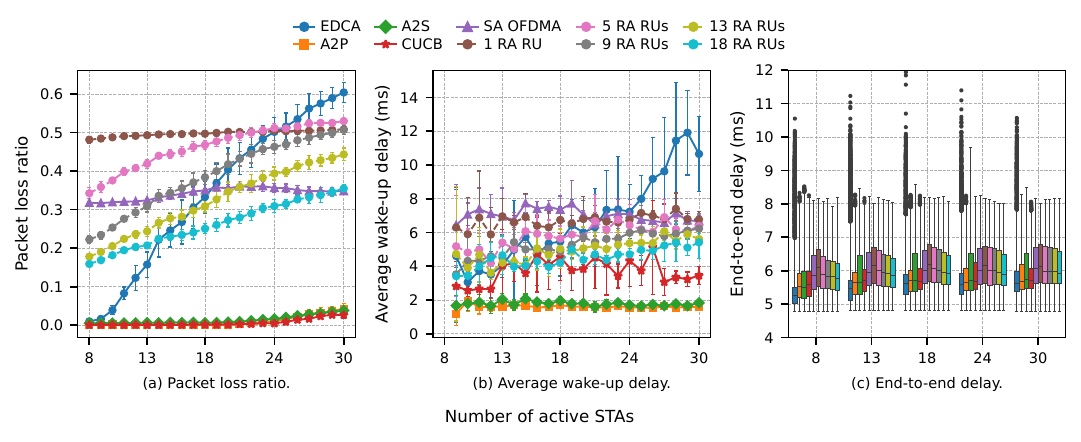}
  \vspace{-0.7cm}
  \caption{Conferencing scenario versus the number of active \glspl{sta} for each
  scheme.}
  \label{fig:conf_results}
\end{figure*}

Having identified the \gls{cmab} configuration, we compare it with the baseline
access schemes. \Cref{fig:conf_results} presents the packet-loss ratio, wake-up
delay, and end-to-end delay as the number of active \glspl{sta} varies.

The packet-loss ratio provides the clearest distinction between the schemes
(cf. \cref{fig:conf_results}(a)). \Gls{a2p}, \gls{a2s}, and \gls{cucb} maintain
zero packet loss up to \num{23} active \glspl{sta} and remain below \num{0.05}
at \num{30}, where their packet-loss ratios are \num{0.043}, \num{0.038}, and
\num{0.027}, respectively. In contrast, \gls{edca} degrades rapidly with load,
reaching \num{0.60} at \num{30} active \glspl{sta} as contention increases.
\Gls{sa-ofdma} maintains a nearly constant loss of approximately
\numrange{0.32}{0.36} because its fixed poll-and-allocate cycle assigns
\glspl{ru} without accounting for \gls{sta} activity. Increasing the number of
\gls{ra} \glspl{ru} in \gls{uora} reduces this inefficiency by shifting
resources toward \glspl{sta} with pending traffic and distributing contention
across more \glspl{ru}. However, this advantage diminishes at high load, where
\gls{uora} and \gls{sa-ofdma} both reach a packet-loss ratio of approximately
\num{0.35}.

The wake-up delay follows the same ordering
(cf. \cref{fig:conf_results}(b)). \Gls{a2p} and \gls{a2s} remain below
\qty{2.5}{\milli\second} across the evaluated load range, while \gls{cucb}
reaches \qty{3.45}{\milli\second} at \num{30} active \glspl{sta}. By limiting
contention from already active \glspl{sta}, \gls{a2p} and \gls{a2s} make
access for newly active \glspl{sta} largely independent of the offered load.
In contrast, the wake-up delay of \gls{edca} increases to
\qty{10.66}{\milli\second} as more \glspl{sta} contend for channel access,
whereas additional \gls{ra} \glspl{ru} reduce the wake-up delay of
\gls{uora}. The round-trip delay in \cref{fig:conf_results}(c) is dominated
by the \gls{dl} tail because the \gls{ul} audio delay is bounded by
\qty{5}{\milli\second}. \Gls{edca} additionally exposes the \gls{dl} broadcast
to contention with \gls{ul} traffic, whereas the scheduled schemes avoid this
contention.

\begin{figure}[!t]
  \centering
  \includegraphics[width=0.8\linewidth]{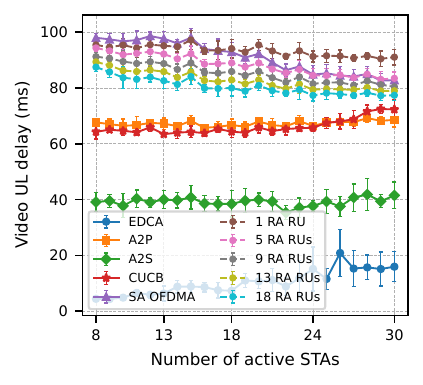}
  \caption{Video uplink delay versus the number of active \glspl{sta} for each
  scheme.}
  \label{fig:conf_video_delay}
  \vspace{-.3cm}
\end{figure}

The video flow behaves differently because it originates from a single
video-class (AC\_VI) \gls{sta}. \Cref{fig:conf_video_delay} shows that
\gls{edca} achieves the lowest video \gls{ul} delay,
\qty{16}{\milli\second} at \num{30} active \glspl{sta}, because the camera
experiences no in-class contention and can use the full channel bandwidth.
The scheduled schemes instead defer each I-frame until the next \gls{ul}
trigger and transmit it on a single 26-tone \gls{ru}, increasing the delay from
\qtyrange{41}{72}{\milli\second}. Nevertheless, fewer than
\qty{1}{\percent} of video packets remain undelivered for all schemes.

Overall, \gls{a2p}, \gls{a2s}, and \gls{cucb} perform best for the
conferencing workload because they direct polling resources toward active
\glspl{sta}. They are the only schemes that keep the audio packet-loss ratio
below \num{0.05} at \num{30} active \glspl{sta} while providing wake-up delays
of only a few milliseconds. \Gls{edca} is competitive only at low load,
whereas \gls{sa-ofdma} wastes scheduled resources on idle \glspl{sta}.
\Gls{uora} reduces this inefficiency with additional \gls{ra} \glspl{ru}, but
does not match schemes that explicitly track \gls{sta} activity.

\subsection{Industrial environment}
\begin{figure*}[!t]
    \centering
    \includegraphics[width=\textwidth]{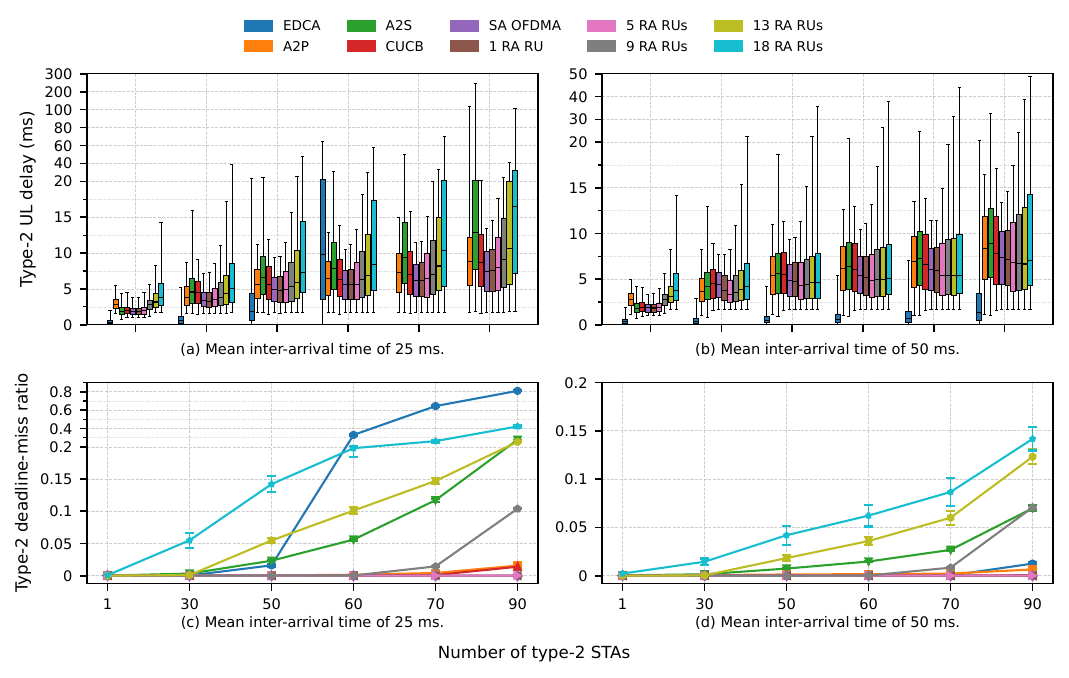}
    \caption{Type-2 uplink delay (a),(b) and packet loss ratio (c),(d) versus
        the number of Type-2 \glspl{sta}.
       }
    \label{fig:iiot_timecritical}
\end{figure*}

\subsubsection{\glsentryshort{cmab} reward and hyperparameter selection}
\label{sec:iiot_cmab}

We repeat the selection procedure of \Cref{sec:conf_cmab}, selecting the
best-performing \gls{cmab} algorithm together with its associated reward
function and hyperparameter based on the Type-2 deadline miss ratio and packet
delivery ratio. The corresponding selection figures are omitted due to space
constraints. The packet delivery ratio prevents configurations from achieving a
low deadline miss ratio by starving Type-2 traffic. The \nth{99}-percentile
delay of Type-2 traffic and Type-1 throughput are used as coexistence
tie-breakers. Since Type-1 traffic is always backlogged, the binary reward tends
to allocate airtime to Type-1 traffic at the expense of the sporadic Type-2
class, resulting in substantially poorer Type-2 performance than the
fairness-weighted reward. With the binary reward, \gls{cucb} loses about
\num{0.97} of the Type-2 traffic at \num{90} stations, whereas Thompson sampling
loses \num{0.85}. Among the fairness-weighted configurations, \gls{cucb}
achieves the lowest Type-2 deadline miss ratio, which decreases as the
exploration constant increases, from \num{0.010} at $c=\sqrt{2}$ to \num{0.0059}
at $c=2$ and \num{0.0046} at $c=2.5$ (mean over \numrange{30}{90} stations at a
\qty{25}{\milli\second} load). We therefore use \gls{cucb} with the
fairness-weighted reward and $c=2.5$ for the remainder of the \gls{iiot}
evaluation.

\subsubsection{Scheme comparison}
\label{sec:iiot_compare}

We first examine the Type-2 \gls{ul} delay, which determines whether the
time-sensitive devices meet the \qty{20}{\milli\second} deadline as their
population grows. Figures~\ref{fig:iiot_timecritical}(a)
and~\ref{fig:iiot_timecritical}(b) correspond to mean Type-2 inter-arrival
times of \qty{25}{\milli\second} and \qty{50}{\milli\second}, respectively.
Under the heavy load, the median delay at \num{90} \glspl{sta} remains below
approximately \qty{17}{\milli\second} for all schemes except \gls{edca},
whose delay exceeds the plotted range at \num{70} and \num{90} \glspl{sta}.
The schemes separate in the upper tail. \gls{sa-ofdma} attains the shortest
tail, with a \nth{99}-percentile delay of \qty{13.4}{\milli\second} at
\num{90} \glspl{sta}, followed by \gls{cucb} with \qty{20.8}{\milli\second}.
\gls{a2p} reaches \qty{115.9}{\milli\second}, since a \gls{sta} not polled in
the current \gls{bsrp} cycle must wait for a subsequent one. \gls{a2s}
removes the polling step yet reaches \qty{245.4}{\milli\second}, since
without buffer-status information transmission opportunities are also
allocated to idle \glspl{sta}. For \gls{uora}, the tail grows from
\qty{14.6}{\milli\second} with one \gls{ra} \gls{ru} to
\qty{106.6}{\milli\second} with \num{18}, as unsuccessful \glspl{sta} contend
again after back-off; at high \gls{sta} counts, the loss of deterministic
scheduled service outweighs the additional \gls{ra} opportunities.

Figures~\ref{fig:iiot_timecritical}(c) and~\ref{fig:iiot_timecritical}(d)
show the Type-2 deadline-miss ratio. Under the heavy load at \num{90}
\glspl{sta} (\cref{fig:iiot_timecritical}c), \gls{sa-ofdma} and \gls{uora}
with a single \gls{ra} \gls{ru} remain below \num{0.001}, \gls{cucb} and
\gls{a2p} reach \num{0.014} and \num{0.016}, and \gls{a2s} reaches
\num{0.28}. The \gls{edca} ratio rises from nearly zero at \num{50}
\glspl{sta} to \num{0.81} at \num{90}, as uncoordinated access leads to
growing contention, retransmissions, and back-off delays. The \gls{uora}
ratio increases with the number of \gls{ra} \glspl{ru}, mirroring its delay
tail. Under the lighter load (\cref{fig:iiot_timecritical}d), the halved
Type-2 offered load narrows the spread in deadline-miss ratios to at most
\num{0.14}. \gls{edca} then achieves a ratio of \num{0.013} and the lowest
median delay, since a \gls{sta} gains access after a short back-off rather
than waiting for a scheduled transmission opportunity.

\begin{figure*}[!t]
  \centering
  \includegraphics[width=\textwidth]{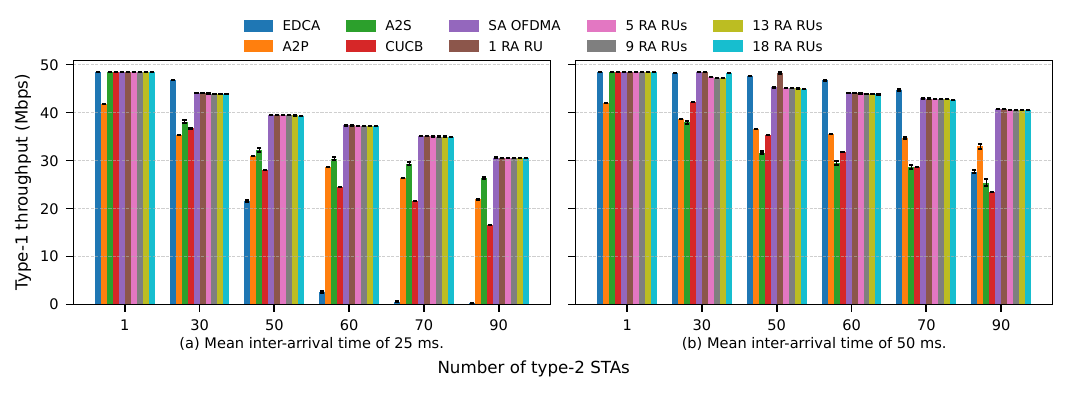}
  \caption{Type-1 background \gls{cbr} throughput versus the number of Type-2
      stations.
    }
  \label{fig:iiot_t1tput}
\end{figure*}

Meeting Type-2 deadlines should not come at the expense of the always-on
Type-1 background traffic, whose \gls{cbr} throughput is shown in
\cref{fig:iiot_t1tput}. At \num{90} \glspl{sta} under the heavy load
(\cref{fig:iiot_t1tput}a), \gls{sa-ofdma} and \gls{uora} with a low number of
\gls{ra} \glspl{ru} retain the highest Type-1 throughput of
\qty{30.6}{\mega\bit\per\second}, whereas \gls{edca} falls from the full
background rate of approximately \qty{48}{\mega\bit\per\second} at low
\gls{sta} counts to \qty{0.3}{\mega\bit\per\second}, consistent with its
Type-2 degradation. \gls{a2p} and \gls{cucb} deliver only
\qty{21.9}{\mega\bit\per\second} and \qty{16.6}{\mega\bit\per\second}; for
\gls{cucb}, the fairness-weighted reward spreads airtime across \glspl{sta},
limiting the share of the always-backlogged Type-1 traffic. The \gls{a2s}
throughput changes little between the two loads
(\qtylist{26.4;25.4}{\mega\bit\per\second}), consistent with its blind
allocation, which cannot recover Type-1 capacity when the Type-2 load
decreases, whereas \gls{sa-ofdma} recovers to
\qty{40.7}{\mega\bit\per\second} under the lighter load.

Overall, under the heavy Type-2 load, \gls{sa-ofdma} and \gls{uora} with a
low number of \gls{ra} \glspl{ru} provide the best balance between the two
traffic classes, combining near-zero Type-2 deadline-miss ratios with the
highest Type-1 throughput. \gls{a2p} and \gls{cucb} also keep Type-2
deadline-miss ratios low but sacrifice Type-1 throughput, increasing the
number of \gls{ra} \glspl{ru} in \gls{uora} substantially worsens Type-2
deadline performance, and \gls{edca} fails on both classes at high \gls{sta}
counts. Under the lighter load, these differences become less pronounced and
\gls{edca} no longer exhibits its throughput collapse.

\subsection{\gls{xr} scenario}

\begin{figure}[!t]
\centering
\includegraphics[width=\columnwidth]{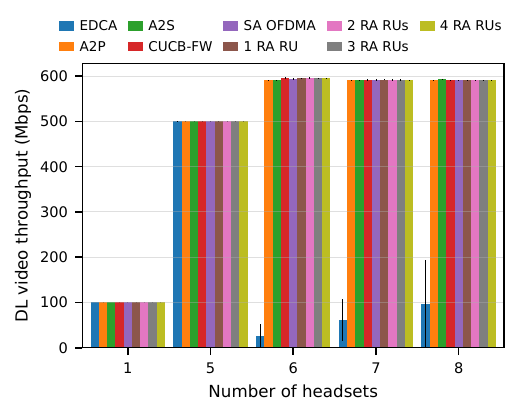}
\caption{\gls{dl} video throughput as a function of number of headsets
    (a.k.a., \glspl{sta}).
    }
\label{fig:xr-video-throughput}
\end{figure}

\begin{figure*}[!t]
\centering
\includegraphics[width=\textwidth]{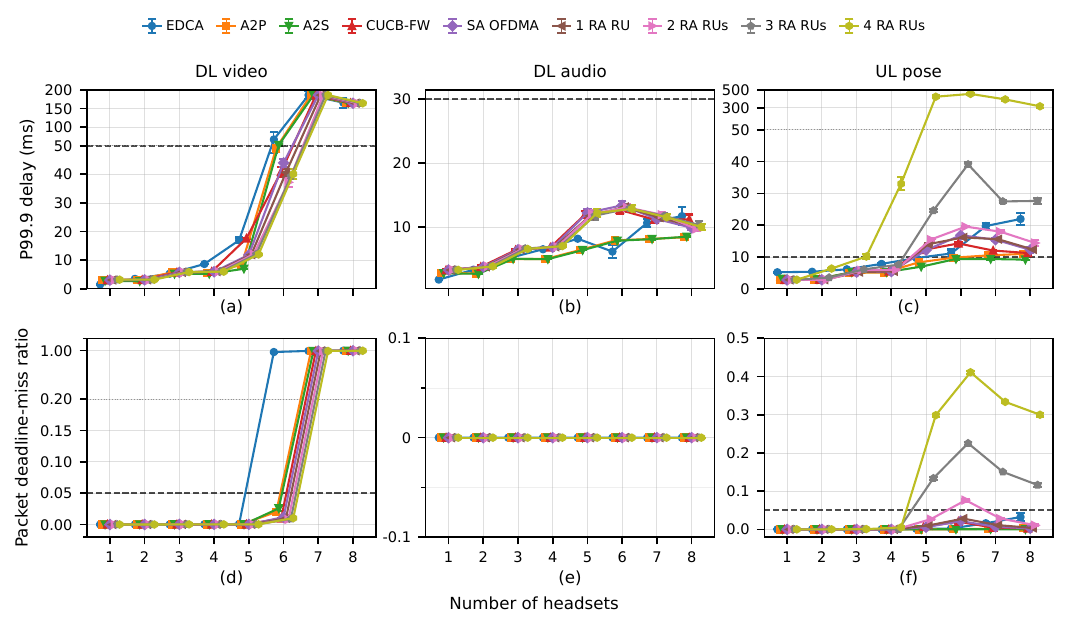}
\caption{Tail delay (top) and deadline miss ratio (bottom) for \gls{dl} video
    ((a),(d)), \gls{dl} audio ((b),(e)) and \gls{ul} pose ((c),(f)) against the
    number of \gls{xr} headsets (a.k.a., \glspl{sta}). Curves are offset
    slightly along the horizontal axis so that overlapping markers remain
    distinguishable, all values correspond to the integer \gls{sta} counts.
    }
\label{fig:xr-p999-loss}
\end{figure*}

\subsubsection{\glsentryshort{cmab} reward and hyperparameter selection}
\label{sec:xr_cmab}

We select the \gls{cmab} algorithm, reward function, and hyperparameter
primarily by the \gls{ul} head-tracking (pose) deadline-miss ratio, using the 
pose and \gls{dl} video \nthd{99.9}-percentile delays as tie-breakers. We omit
the corresponding figures due to space constraints. Since the pose deadline-miss
ratio is aggregated over all \glspl{sta}, a configuration can achieve a low
ratio while persistently underserving a single \gls{sta}. We therefore use the
minimum per-\gls{sta} pose delivery ratio to expunge such configurations. The
\gls{dl} video deadline-miss ratio is not used as a selection criterion, as it
varies by less than \num{0.006} across the evaluated configurations.

The reward function has the largest effect on the stringent
deadline requirement of the pose traffic. The binary
reward assigns the same value to every positive \gls{bsr}, so it neither
distinguishes among \glspl{sta} with queued pose traffic nor penalizes
repeated selection of the same \gls{sta}. The fairness-weighted reward
instead combines the reported buffer sizes with a penalty on previous
selections, shifting service toward less frequently served \glspl{sta}. With
\gls{cucb} at \num{5} \glspl{sta}, the fairness-weighted reward reduces the
pose deadline-miss ratio from \num{0.138} to \num{0.0067} and the
\nthd{99.9}-percentile delay from \qty{189.2}{\milli\second} to
\qty{12.2}{\milli\second}, and at \num{6} \glspl{sta} it raises the minimum
per-\gls{sta} pose delivery ratio from \num{0.949} to \num{0.9997}.
We therefore use \gls{cucb} with the fairness-weighted reward and $c=2.5$ for
the remainder of the \gls{xr} evaluation.

\subsubsection{Scheme comparison}
\label{sec:xr_compare}

The \gls{dl} video traffic dominates the offered load, with each \gls{sta}
receiving an offered video rate of \qty{100}{\mega\bit\per\second}
(cf. \cref{fig:xr-video-throughput}). Up to \num{5} \glspl{sta}, all schemes
deliver the full offered load well within the \qty{50}{\milli\second}
deadline (cf. \cref{fig:xr-p999-loss}a). The schemes separate at \num{6}
\glspl{sta}. The \gls{ofdma}-based schemes keep the video deadline-miss ratio
at or below \num{0.026} and deliver nearly the full \qty{600}{\Mbps} offered
load, whereas under \gls{edca} \qty{97.4}{\percent} of video packets miss
their deadline and only \qty{26.5}{\Mbps} is delivered
(cf. \cref{fig:xr-p999-loss}d). The \gls{ap} holds a saturated video queue
but contends as a single entity against the \glspl{sta}, so its channel
access opportunities do not scale with the video load. \gls{ofdma} avoids
this bottleneck by serving multiple \glspl{sta} in a single multi-user
transmission. Beyond \num{6} \glspl{sta}, the \gls{ofdma}-based schemes
saturate at approximately \qty{591}{\Mbps}, largely independent of the scheme
and the number of \gls{ra} \glspl{ru}.

Although the \gls{ofdma}-based schemes satisfy the video deadline-miss
requirement at \num{6} \glspl{sta}, their delay tails approach the deadline.
\gls{a2s} and \gls{a2p} exhibit the longest worst-station
\nthd{99.9}-percentile delays, at \qty{51.8}{\milli\second} and
\qty{49.2}{\milli\second}, compared with \qty{37.2}{\milli\second} for
\gls{uora} with two \gls{ra} \glspl{ru}, since both reserve \gls{ul}
opportunities at a fixed \gls{ari} and thereby reduce the airtime available
for \gls{dl} video transmissions. The deadline-miss ratio counts generated
packets that fail the deadline, whereas the \nthd{99.9} percentile
characterises the extreme tail of the delivered-packet delays. By the former
criterion, \num{6} \glspl{sta} remain feasible, although performance is close
to the video latency limit (cf. \cref{fig:xr-p999-loss}a,d).

The \gls{dl} audio flow does not constrain the supported number of
\glspl{sta} (cf. \cref{fig:xr-p999-loss}b,e). Its deadline-miss ratio is zero
for all schemes and \gls{sta} counts, and its worst-station
\nthd{99.9}-percentile delay peaks at \qty{13.4}{\milli\second}, well below
the \qty{30}{\milli\second} deadline. The audio queue is served ahead of the
saturated video queue and contributes only \qty{1.85}{\Mbps} per \gls{sta},
so it drains unconstrained by the video traffic.

The \gls{ul} pose flow depends most strongly on the scheme
(cf. \cref{fig:xr-p999-loss}c,f). At \num{5} \glspl{sta}, the pose
deadline-miss ratio grows monotonically with the number of \gls{ra}
\glspl{ru}, from \num{0.0058} under \gls{sa-ofdma} to \num{0.299} with four
\gls{ra} \glspl{ru}, where the \nthd{99.9}-percentile delay reaches
\qty{424.3}{\milli\second}. As in the \gls{iiot} scenario, replacing
scheduled service with contention degrades deadline performance despite the
additional \gls{ra} opportunities. Among the \gls{ofdma}-based schemes,
\gls{a2p} and \gls{a2s} achieve the lowest pose deadline-miss ratios,
\num{0.0001} at \num{5} \glspl{sta} and \num{0.0005} at \num{6}, compared
with \num{0.0203} for \gls{cucb} and \num{0.0217} for \gls{sa-ofdma}. Both
maintain a poll list of recently active \glspl{sta}, which a \gls{sta}
re-enters at the head upon new activity and leaves after
\qty{40.96}{\milli\second} without service, whereas \gls{sa-ofdma} and
\gls{cucb} rotate through all associated \glspl{sta}. Since pose packets
arrive at independent instants within each \qty{11.11}{\milli\second} frame
interval, a \gls{sta} whose packet arrives just after its turn must wait for
the rotation to return to it. The activity-based poll list reduces this
dependence on the rotation order. At \num{6} \glspl{sta}, \gls{uora} with two
or more \gls{ra} \glspl{ru} exceeds the \num{0.05} pose deadline-miss
threshold, while \gls{edca} maintains a ratio of \num{0.0023}. The short pose
packets originate at the \glspl{sta} and contend directly for channel access,
so their delivery does not depend on the \gls{ap} draining its saturated
video queue.

The choice of scheme therefore trades \gls{dl} video latency against \gls{ul}
pose reliability. \gls{a2p} and \gls{a2s} achieve the lowest pose
deadline-miss ratios but the longest video delay tails, while \gls{sa-ofdma}
and \gls{cucb} show the opposite balance. Increasing the number of \gls{ra}
\glspl{ru} reduces scheduled \gls{ul} service and pushes the pose flow past
its deadline-miss threshold from two \gls{ra} \glspl{ru} onward at \num{6}
\glspl{sta}. Nevertheless, all \gls{ofdma}-based schemes support \num{6}
\glspl{sta} within the video deadline-miss requirement, whereas \gls{edca}
fails to sustain the video traffic at this load.

\subsection{Takeaways}

\begin{table*}[!t]
    \centering
    \caption{Summary of Scheme Suitability per Scenario}
    \label{tab:summary}
    \renewcommand{\arraystretch}{1.4}
    \begin{tabular}{l p{4.6cm} p{4.6cm} p{4.6cm}}
        \hline
        \textbf{Scheme} & \textbf{Conferencing} & \textbf{Industrial
        \glsxtrshort{iot}} & \textbf{\glsxtrshort{xr}} \\
        \hline
        \gls{edca} &
        Competitive only at low load & 
        Lowest median delay under light load. Collapses beyond \num{60}
        \glspl{sta} under heavy load &
        Video unsustainable beyond \num{5} \glspl{sta}. Pose traffic unaffected \\
        \gls{sa-ofdma} &
        Wastes polls on idle \glspl{sta}. Load-invariant deadline-miss ratio of
        ${\approx}\num{0.35}$ &
        \textbf{Best balance}: near-zero Type-2 deadline misses with the
        highest Type-1 throughput &
        Short video delay tails, but elevated pose deadline misses \\
        \gls{uora} &
        More \gls{ra} \glspl{ru} reduce deadline misses, yet short of the
        activity-tracking schemes &
        \textbf{Best, matches \glsxtrshort{sa-ofdma} with one \gls{ra}
        \gls{ru}}. Degrades as \gls{ra} \glspl{ru} increase &
        Pose deadline misses exceed the threshold from two \gls{ra} \glspl{ru} onward \\
        \gls{a2p} &
        \textbf{Best}. Audio loss below \num{0.05} with millisecond wake-up
        delays &
        Low Type-2 deadline misses at reduced Type-1 throughput &
        \textbf{Best}. Lowest pose deadline misses. Video tail at the deadline \\
        \gls{a2s} &
        \textbf{Best}, as \gls{a2p}, without the polling overhead &
        High Type-2 deadline misses due to buffer-blind allocation &
        \textbf{Best}. Lowest pose deadline misses. Longest video tail \\
        \gls{cmab} (\gls{cucb}) &
        \textbf{Best}, with the binary reward &
        Low Type-2 deadline misses, but lowest Type-1 throughput &
        Short video delay tails, but elevated pose deadline misses \\
        \hline
    \end{tabular}
\end{table*}

When traffic is sporadic and only a subset of \glspl{sta} is active at a
time, as in the conferencing scenario and for the \gls{xr} pose flow, schemes
that use information about \gls{sta} activity or buffer state (\gls{a2p},
\gls{a2s}, and \gls{cmab}) perform best because they direct transmission
opportunities to \glspl{sta} with pending traffic. When traffic demand is
persistent, as in the industrial scenario, where the Type-1 background keeps
the network persistently loaded, and for the \gls{xr} video flow, scheduled
access performs better, with \gls{sa-ofdma} and \gls{uora} with a single
\gls{ra} \gls{ru} providing the strongest performance under the evaluated
conditions. Contention-based access remains viable only at relatively light
load. \Cref{tab:summary} summarizes the suitability of each scheme per
scenario.

\section{Conclusion}
\label{sec:conclusion}

In this paper, we examined which buffer-status polling scheme is best suited to
different time-sensitive Wi-Fi workloads. We compared \gls{edca},
\gls{sa-ofdma}, \gls{uora}, \gls{a2p}, \gls{a2s}, and \gls{cmab}-based polling
under similar traffic and channel conditions within three environments. For each
scheme and scenario, we selected the best-performing configuration among the
evaluated settings. Our results show that no single scheme performs best across
all three environments. A consistent pattern nevertheless emerged. Activity- and
learning-based polling (\gls{a2p}, \gls{a2s}, and \gls{cmab}) prevails when
traffic is sporadic and \gls{sta} activity varies, whereas fixed scheduled
service (\gls{sa-ofdma}, or \gls{uora} with few \gls{ra} \glspl{ru}) prevails
when offered load demand is persistent, and contention-based access remains
viable only at highly sporadic light load.

In the conferencing environment, activity- and learning-based polling
maintained low wake-up delay for newly active talkers as the offered load
increased, whereas \gls{edca} remained competitive only under light load. In
the \gls{iiot} environment, scheduled service provided the strongest deadline
performance, with \gls{sa-ofdma} keeping the Type-2 deadline-miss ratio below
\num{0.001} at \num{90} stations where \gls{edca} reached \num{0.81}. In the
\gls{xr} environment, every \gls{ofdma}-based scheme sustained the dominant
\gls{dl} video flow at six headsets, so the schemes separated primarily on
the tighter \gls{ul} pose flow, where the activity-based schemes achieved the
lowest deadline-miss ratios. Across the evaluated \gls{cmab} configurations,
the reward signal had a larger effect on performance than the choice of
learning algorithm, and the preferred reward depended on the traffic
characteristics of the environment

These conclusions hold under the assumptions of our setup. Each experiment
models a single \gls{bss} with no overlapping \glspl{bss} and no hidden
terminals, and the transmit power is set so that packet loss from channel
errors is negligible. Whether the ranking of the schemes persists under
inter-\gls{bss} contention and channel-induced loss remains to be verified.
Each scheme also runs the fixed configuration that performed best in an
offline sweep, whereas a deployed network sees its workload change. Selecting
the polling scheme and its parameters online is therefore itself a learning
problem, and our \gls{cmab} results indicate that the design of the reward
signal, more than the choice of learning algorithm, will determine whether
such a selector works.


\printbibliography[title={References}]

\glsresetall
\appendix[Combinatorial Multi-Armed Bandit Algorithms]
\label[appendix]{appen:cmab}

We formulate the selection of \glspl{sta} to poll as a stochastic
\glsxtrfull{cmab} problem~\cite{Chen2013CUCB}. Let $\mathcal{N}=\{1,\dots,N\}$
denote the set of associated \glspl{sta}, with each \gls{sta} corresponding to a
base arm. At polling round $t \in \{1,\dots,T\}$, where $T$ is the total number
of polling rounds, the \gls{ap} selects a super-arm
$\mathcal{S}_t\subseteq\mathcal{N}$ with cardinality
\begin{equation}
    |\mathcal{S}_t|
    =
    K_t
    =
    \min(N,R_t),
\end{equation}
where $R_t$ is the number of \glspl{ru} available in round $t$. Each selected
arm $i\in\mathcal{S}_t$ yields a reward $r_{i,t}\in[0,1]$ derived from the
buffer-size field contained in the QoS~Null response to the \gls{bsrp} \gls{tf}
(\cref{subsec:saofdma}).

Let $\mu_{i,t}=\mathbb{E}[r_{i,t}]$ denote the expected reward of arm $i$ at
round $t$, and define the expected reward of a super-arm as
\begin{equation}
    \mu_t(\mathcal{S})
    =
    \sum_{i\in\mathcal{S}}\mu_{i,t}.
\end{equation}
The optimal feasible super-arm at round $t$ is
\begin{equation}
    \mathcal{S}_t^\star
    =
    \operatorname*{arg\,max}_{\substack{\mathcal{S}\subseteq\mathcal{N}\\
    |\mathcal{S}|=K_t}}
    \mu_t(\mathcal{S}).
\end{equation}
The objective is to minimise the cumulative pseudo-regret
\begin{equation}
    \mathcal{R}_T
    =
    \mathbb{E}\!\left[
    \sum_{t=1}^{T}
    \left(
        \mu_t(\mathcal{S}_t^\star)
        -
        \mu_t(\mathcal{S}_t)
    \right)
    \right],
    \label{eq:regret}
\end{equation}
which measures the expected reward lost relative to selecting the best feasible
super-arm at each polling round.

\subsubsection{Reward signals}
\label{subsec:cmab-reward}

The reward signal maps the \gls{bsr} reported by each selected \gls{sta} to a
scalar value used by the active bandit policy. Let $b_{i,t}$ denote the
buffered-byte estimate reported by \gls{sta} $i$ at round $t$. We consider two
reward signals that differ in how much information from the report and the
polling history is used in apportioning the reward.

\paragraph{Binary reward.}
The binary reward indicates only whether the selected \gls{sta} reports pending
data:
\begin{equation}
    r_{i,t}^{\mathrm{bin}}
    =
    1,~\mathrm{if}~b_{i,t}>0.
    \label{eq:reward-binary}
\end{equation}
A positive buffer status therefore produces a reward of one regardless of the
reported queue size. The reward also contains no information about how
frequently the \gls{sta} has previously been selected. Consequently, two
\glspl{sta} with positive buffer reports receive the same reward even when their
queue occupancies or polling histories differ. The binary signal therefore
identifies whether polling a \gls{sta} reveals pending traffic, but provides no
additional discrimination among positive reports.

\paragraph{Fairness-weighted reward.}
The fairness-weighted reward incorporates both the reported buffer size and
the polling history of the \gls{sta}. Let $n_i$ denote the number of times arm
$i$ has been selected up to the current round, and let $B_{\max}$ denote the
buffer-size normalisation constant. We define
\begin{equation}
    r_{i,t}^{\mathrm{fw}}
    =
    \underbrace{
    \min\!\left(1,\frac{b_{i,t}}{B_{\max}}\right)
    }_{\text{normalised buffer occupancy}}
    \,
    \underbrace{
    \frac{1}{1+n_i/N}
    }_{\text{polling-history penalty}}.
    \label{eq:reward-fairness}
\end{equation}
The first factor increases with the reported backlog and is bounded by one,
allowing \glspl{sta} with more queued data to receive a larger reward than those
with low or empty queues. The second factor decreases as a \gls{sta} is
selected more frequently. Repeatedly polled \glspl{sta} therefore receive a
progressively larger penalty, increasing the relative reward of less frequently
selected \glspl{sta}. The combined reward balances the reported demand against
the previous polling frequency while remaining bounded in $[0,1]$.

Both reward signals are evaluated online from the current buffer report, with the
fairness-weighted reward additionally changing with the evolving pull count
$n_i$. We consequently use the \gls{cmab} algorithms as online
\gls{sta}-selection policies without assuming a stationary reward distribution.

\subsubsection{\Gls{cucb}}
\label{sec:cucb}

For each \gls{sta} $i \in \mathcal{N}$, the \gls{ap} maintains two statistics;
the pull count $n_i$, recording the number of polling rounds in which \gls{sta}
$i$ was included in the selected super-arm, and the empirical mean reward
$\hat{\mu}_i$. \Glspl{sta} are selected according to an upper-confidence
estimate of their expected reward. At round $t$, $i$ is assigned the score
\begin{equation}
    U_{i,t}
    =
    \hat{\mu}_i
    +
    c\sqrt{\frac{2\ln t}{n_i}},
    \label{eq:ucb}
\end{equation}
where $c>0$ is the exploration constant. The first term favours \glspl{sta} with high
historical reward, whereas the second term increases the score of those
infrequently polled, thereby promoting exploration.  The \gls{ap} constructs the
super-arm by selecting the $K_t$ arms with the largest \gls{ucb} scores,
$
    \mathcal{S}_t
    =
    \operatorname*{top}_{K_t}
    \bigl\{
        U_{i,t}
    \bigr\}_{i\in\mathcal{N}}.
    \label{eq:cucb-selection}
$
\Glspl{sta} with $n_i=0$ are assigned $U_{i,t}=+\infty$ to guarantee that every
\gls{sta} is polled at least once during the initial exploration phase. After
the transmissions of the selected \glspl{sta}, the \gls{ap} observes the
individual rewards $\{r_{i,t}\}_{i\in\mathcal{S}_t}$ and updates the
corresponding statistics for each selected arm accordingly;
$
    \hat{\mu}_i
    \leftarrow
    \hat{\mu}_i
    + \frac{1}{n_i}\bigl( r_{i,t} - \hat{\mu}_i \bigr)
$
$
    n_i \leftarrow n_i + 1,
$
. \glspl{sta} outside $\mathcal{S}_t$ retain their previous estimates.

\subsubsection{\Gls{cts}}
\label{subsec:cts}

\gls{cts} models the expected reward of each \gls{sta} using a Bayesian
posterior distribution~\cite{WangChen2018CTS}. For each \gls{sta} $i \in
\mathcal{N}$, the \gls{ap} maintains a Beta posterior
$\mathrm{Beta}(\alpha_i,\beta_i)$. At polling round $t$, the \gls{ap} independently
draws a posterior sample
$
    \tilde{\theta}_{i,t}
    \sim
    \mathrm{Beta}(\alpha_i,\beta_i),
    \label{eq:ts-sample}
$
then selects the $K_t$ \glspl{sta} with the largest sampled reward
estimates,
$
    \mathcal{S}_t
    =
    \operatorname{top\text{-}}K_t
    \bigl\{
        \tilde{\theta}_{i,t}
    \bigr\}_{i\in\mathcal{N}}.
$
Samples from the Beta posterior are generated using two independent Gamma
variates,
$
    G_1\sim\Gamma(\alpha_i,1)
$
and
$
    G_2\sim\Gamma(\beta_i,1),
$
according to
\begin{equation}
    \tilde{\theta}_{i,t}
    =
    \frac{G_1}{G_1+G_2}.
\end{equation}
The \gls{ap} observes the rewards $r_{i,t}\in[0,1]$ after the transmissions of
the scheduled \glspl{sta} and updates their posterior parameters as
$
    \alpha_i
    \leftarrow
    \alpha_i + r_{i,t},
    \qquad
    \beta_i
    \leftarrow
    \beta_i + 1-r_{i,t},
$
which naturally extends the standard Bernoulli Beta-Binomial update to
continuous rewards in the interval $[0,1]$.

\subsubsection{EXP3.M}
\label{sec:exp3m}

\gls{exp3m}~\cite{Uchiya2010EXP3M} extends the classical \glsfmtshort{exp3}
algorithm~\cite{Auer2002EXP3} to combinatorial \gls{sta} selection, where $K_t$
\glspl{sta} are selected simultaneously under adversarial or non-stationary
reward processes. For each \gls{sta} $i \in \mathcal{N}$, the \gls{ap} maintains
a non-negative weight $w_{i,t}$, initialised to $w_{i,1}=1$. At polling round
$t$, these weights are mapped to inclusion probabilities via a mixture of
exploitation and exploration:
\begin{equation}
    \bar{p}_{i,t}
    =
    (1-\gamma)\frac{w_{i,t}}{\sum_{j \in \mathcal{N}} w_{j,t}}
    +
    \frac{\gamma}{|\mathcal{N}|},
    \label{eq:exp3m-prob-raw}
\end{equation}
where $\gamma \in (0,1]$ controls the exploration–exploitation trade-off. The
vector $\bar{p}_t$ is then normalised to enforce a valid $K_t$-subset selection,
since direct scaling by $K_t$ may lead to $\bar{p}_{i,t} > 1$ when the weight
distribution is highly skewed. Specifically, a threshold $\alpha_t > 0$ is
computed such that
\begin{equation}
    \alpha_t
    \sum_{i : \bar{p}_{i,t} < \alpha_t}
    \frac{w_{i,t}}{\sum_{j \in \mathcal{N}} w_{j,t}}
    +
    \sum_{i : \bar{p}_{i,t} \ge \alpha_t} 1
    =
    K_t,
    \label{eq:exp3m-cap}
\end{equation}
The resulting inclusion probabilities are given by
\begin{equation}
    p_{i,t}
    =
    \begin{cases}
    1, & \bar{p}_{i,t} \ge \alpha_t,\\[4pt]
    \alpha_t \dfrac{w_{i,t}}{\sum_{j \in \mathcal{N}} w_{j,t}}
    (1-\gamma)
    +
    \frac{\gamma K_t}{|\mathcal{N}|}, & \text{otherwise},
    \end{cases}
    \label{eq:exp3m-prob}
\end{equation}
which guarantees $0 \le p_{i,t} \le 1$ and $\sum_{i \in \mathcal{N}} p_{i,t} =
K_t$. The \gls{ap} then samples a super-arm $\mathcal{S}_t$ of size $K_t$
without replacement using weighted reservoir
sampling~\cite{Efraimidis2006WROS}. After transmission, the \gls{ap} observes
rewards $r_{i,t} \in [0,1]$ for $i \in \mathcal{S}_t$ and forms the
importance-weighted estimate
\begin{equation}
    \tilde{r}_{i,t}
    =
\begin{cases}
\frac{r_{i,t}}{p_{i,t}}, & i \in \mathcal{S}_t,\\[6pt]
0, & \text{otherwise},
\end{cases}
\label{eq:exp3m-iw-estimate}
\end{equation}
which yields an unbiased estimator of the true reward under partial
feedback. The weights are then updated multiplicatively as
\begin{equation}
w_{i,t+1}
=
w_{i,t}
\exp\!\left(
\frac{\gamma K_t}{|\mathcal{N}|}\tilde{r}_{i,t}
\right).
\label{eq:exp3m-update}
\end{equation}

\subsubsection{$\varepsilon$-Greedy}
\label{sec:eps-greedy}

The $\varepsilon$-greedy baseline alternates between exploitation and
uniform exploration. At polling round $t$, the \gls{ap} selects the
super-arm
\begin{equation}
    \mathcal{S}_t
    =
    \begin{cases}
    \operatorname{top\text{-}}K_t\{\hat{\mu}_i\},
    & \text{with prob. } 1-\varepsilon,\\[4pt]
    \text{random }K_t\text{-subset of }\mathcal{N},
    & \text{with prob. } \varepsilon,
    \end{cases}
    \label{eq:eps-greedy}
\end{equation}
where $\varepsilon \in [0,1]$ controls the exploration rate. During
exploitation, the \gls{ap} selects the $K_t$ \glspl{sta} with the largest
empirical mean rewards $\hat{\mu}_i$, whereas the exploration branch samples a
subset uniformly at random without replacement using reservoir
sampling~\cite{Vitter1985Reservoir}. Similar to \gls{cucb}, \gls{sta} with
$n_i=0$ are assigned infinite scores during exploitation to guarantee that every
\gls{sta} is explored at least once. After observing the reward
$r_{i,t}\in[0,1]$ from each selected \gls{sta} $i\in\mathcal{S}_t$, the \gls{ap}
updates the pull count and empirical mean reward according to
$
    n_i \leftarrow n_i + 1,
$
$
    \hat{\mu}_i
    \leftarrow
    \hat{\mu}_i
    + \frac{1}{n_i}
    \bigl(r_{i,t} - \hat{\mu}_i\bigr).
    \label{eq:eps-update}
$

\end{document}